\documentclass[11pt]{article}

\usepackage{authblk}  
\usepackage{graphicx} 
\usepackage{epsfig}
\usepackage{dcolumn}  
\usepackage{bm}       
\usepackage{amsmath}
\usepackage{amssymb}

\newcommand{\beq}{\begin{equation}} \newcommand{\eeq}{\end{equation}}
\newcommand{\bea}{\begin{eqnarray}} \newcommand{\eea}{\end{eqnarray}}

  \newcommand
{\Romannumeral}[1]{\uppercase\expandafter{\romannumeral#1}}

\newcommand{\be}{\begin{enumerate}} \newcommand{\ee}{\end{enumerate}}
\newcommand{\bi}{\begin{itemize}} \newcommand{\ei}{\end{itemize}}
\newcommand{\ba}{\begin{array}} \newcommand{\ea}{\end{array}}
\newcommand{\bc}{\begin{center}} \newcommand{\ec}{\end{center}}
\newcommand{\bt}{\begin{tabular}} \newcommand{\et}{\end{tabular}}

\def\lsim{\mathrel{\rlap{\lower4pt\hbox{\hskip1pt$\sim$}}
    \raise1pt\hbox{$<$}}}           
\def\gsim{\mathrel{\rlap{\lower4pt\hbox{\hskip1pt$\sim$}}
    \raise1pt\hbox{$>$}}}           

\newcommand{\Tr}{\mathop{\rm Tr}}           
\newcommand{\tr}{\mathop{\rm tr}}           
\newcommand{\half}{\textstyle {1\over2} \displaystyle}    
\newcommand{\third}{\textstyle {1\over3} \displaystyle}   
\newcommand{\quarter}{\textstyle {1\over4} \displaystyle} 
\newcommand{\eigth}{\textstyle {1\over8} \displaystyle}   
\newcommand{\twoth}{\textstyle {2\over3} \displaystyle}   
\newcommand{\Dslash}{{\hbox{D}\kern-0.6em\raise0.15ex\hbox{/}}} 

\renewcommand{\et}{\eta}

\begin{document}

\setlength{\oddsidemargin}{0cm} \setlength{\baselineskip}{7mm}

\begin{normalsize}\begin{flushright}

January 2013 \\

\end{flushright}\end{normalsize}

\begin{center}
  
\vspace{5pt}

{\Large \bf Inconsistencies from a Running Cosmological Constant}

\vspace{30pt}

{\sl Herbert W. Hamber}
$^{}$\footnote{e-mail address : Herbert.Hamber@aei.mpg.de} 
\\
Max Planck Institute for Gravitational Physics \\
(Albert Einstein Institute) \\
D-14476 Potsdam, Germany\\

\vspace{5pt}

and

\vspace{5pt}

{\sl Reiko Toriumi}
$^{}$\footnote{e-mail address : RToriumi@uci.edu} \\
Department of Physics and Astronomy, \\
University of California, \\
Irvine, California 92697-4575, USA \\

\vspace{10pt}

\end{center}

\begin{center} {\bf ABSTRACT } \end{center}

\noindent

We examine the general issue of whether a scale dependent cosmological 
constant can be consistent with general covariance,
a problem that arises naturally in the treatment of quantum gravitation
where coupling constants generally run as a consequence
of renormalization group effects.
The issue is approached from several points of view, which include
the manifestly covariant functional integral formulation,
covariant continuum perturbation theory about two dimensions,
the lattice formulation of gravity, and the non-local
effective action and effective field equation methods.
In all cases we find that the cosmological constant cannot
run with scale, unless general covariance is explicitly broken
by the regularization procedure.
Our results are expected to have some bearing on current quantum gravity
calculations, but more generally should apply to phenomenological
approaches to the cosmological vacuum energy problem.



\vfill

\pagestyle{empty}

\newpage

\pagestyle{plain}

\section{Introduction}
\label{sec:intro}

\vskip 10pt

It is a common feature of the renormalization group approach
to quantum field theory that coupling constant are generally
scale dependent: they run with momentum scale in
a way that is determined by the Callan-Symanzik beta function
and its nonperturbative extensions.
The physical reasons for the scale dependence of couplings
is generally well understood.
It arises because of the effects of virtual quanta that
either screen (as in QED) or anti-screen (as in QCD) the
fundamental bare charge, with the (anti)screening scale
generally determined by a relevant infrared cutoff.
Gravity itself is not immune from such effects, which arise
from graviton and matter vacuum polarization contributions,
but the issues has been clouded for some time as a consequence
of the well-known perturbative non-renormalizability problem.
Nevertheless, a number of physically relevant results have
been obtained by either applying Wilson's $2+\epsilon$ dimensional
expansion method, or via the 4d lattice formulation of gravity
developed by Regge and Wheeler.  
In either case, definite predictions arise for the scale dependence
of Newton's constant $G$, which are generally consistent
between the two approaches.
In particular, the lattice theory predicts a slow rise in the
gravitational coupling with scale, similar to
the well-known anti-screening effect of non-Abelian gauge theories.
In either case the cosmological constant cannot be made to run
with scale, and it emerges instead naturally as a 
nonperturbative renormalization group invariant scale, 
formally an integration constant of the renormalization group equations.
A key ingredient in both theories is the preservation of local
diffeomorphism invariance, which would otherwise spoil this
last result.

In this paper we address the issue of how general the result is
that the cosmological constant cannot be momentum-scale dependent,
if manifest general covariance is strictly maintained.
To do so, we first examine the question of the dependence of the 
renormalization group equations on the bare cosmological constant.
Within the manifestly covariant functional integral approach to 
gravity, it is then easy to show (both in the Euclidean and in
the Lorentzian formulation) that the bare cosmological can be 
entirely scaled out, so that physical invariant correlations cannot depend on it.
The same is found to be true in the lattice formulation of gravity, 
where again the bare cosmological can be scaled out, and thus set equal
to one in units of the ultraviolet cutoff, without any loss
in generality.
One concludes therefore that a running of lambda is meaningless in 
either formulation.

The above conclusions are reinforced by a study of perturbative
gravity in $2+\epsilon$ dimensions.
Here radiative corrections are computed using the background field 
method and dimensional regularization, 
by suitably performing a formal double expansion in $d-2$ and $G$.
In this work we will point out that these results clearly show
that the renormalization of the cosmological constant is
gauge dependent.
Furthermore, this spurious renormalization entirely disappears 
once a suitable rescaling of the metric is performed in order to 
remove the unwanted gauge dependence.

An alternative approach to the problem of the running cosmological
constant is via a set of manifestly
covariant effective field equations, constructed so
as to incorporate the running of Newton's constant $G$, and in
a way that is consistent with the results from the manifestly
covariant functional integral method described earlier.
Again we find that within such a framework it is nearly
impossible to accommodate a scale dependence of the
cosmological constant, for the simple reason that covariant
derivatives of the metric tensor vanish identically.
A similar result is later obtained from a slightly different
approach, centered on an effective action of quantum gravity.

An outline of the paper is as follows.
Sec.~2 discusses scaling properties of the continuum
functional integral for gravity, and in particular
how it transforms under a uniform rescaling of the metric.
Sec.~3 recalls how the problem of the renormalization of the cosmological
constant is resolved in the perturbative treatment
of gravity.
Sec.~4 shows that the bulk of the conclusions of Sec.~2
for the continuum 
are still valid on the lattice, and in particular
the fact that the functional integral does not
depend on the specific value for the bare cosmological constant,
as long as it is positive, and irrespective of the choice 
of functional measure.
The role of the new fundamental nonperturbative scale $\xi$ 
that arises both in the lattice and in the continuum treatment
is emphasized.
Sec.~5 points out similarities between
gravity and non-Abelian gauge theories, and in particular
the important role played, in the nonperturbative 
treatment of QCD, by the dynamically generated length scale,
seen to be related in a simple way to the color vacuum condensate.
Sec.~6 summarizes the key points that lead to the identification,
in the framework of nonperturbative gravity,
of the renormalization-group invariant scale $\xi$
with a gravitational condensate, and thus with
the observed cosmological constant.
Sec.~7 discusses the running of the cosmological constant
from the perspective of a set of manifestly covariant, but
nonlocal, effective field equations. 
Here it is shown that the cosmological constant, in this framework,
cannot run.
Sec.~8 analyzes the same problem by considering the
implications of an effective action formulation, thereby
reaching the same conclusions as in the previous section.
Sec.~9 contains a summary of our results.

\vskip 30pt

\section{Gravitational Functional Integral}

\label{sec:cont}

\vskip 10pt

In this section we recall some basic properties of the functional
integral for gravity, which will lead to the conclusion
that the bare cosmological constant can largely be scaled out of
the problem.
Formally, the Euclidean Feynman path integral for pure
Einstein gravity with a cosmological constant term can be written as
\footnote{Most aspects of the following discussion would remain
unchanged if we were to consider instead the Lorentzian formulation.
For concreteness, we will focus here almost exclusively on the
Euclidean theory.}
\beq
Z \; = \; \int [ d \, g_{\mu\nu} ] \; \exp \, \Bigl \{
- \lambda_0 \, \int d x \, \sqrt g \, + \, 
{ 1 \over 16 \pi \, G } \int d x \, \sqrt g \, R \Bigr \} \;\; .
\label{eq:zcont}
\eeq
The above state sum involves a functional integration over 
all metrics, with measure given by a suitably regularized form of 
\beq
\int [ d \, g_{\mu\nu} ] \; \equiv \; \int \prod_x \; 
\left [ g(x) \right ]^{ \sigma / 2 } \;
\prod_{\mu \ge \nu} \, d g_{\mu \nu} (x) \; ,
\label{eq:measure}
\eeq
as given in Eqs.~(\ref{eq:dw-dewitt}) and (\ref{eq:misner}) below,
and with $\sigma$ some real parameter.
The value of $\sigma$ will play no significant role in the following,
as long as the relevant integrals are known to exist.
For geometries with boundaries, further terms will need to be added
to the action, representing the effects of those boundaries.
Here we will consider the above expression in the absence of such
boundaries.

Let us first focus on some basic scaling properties of the gravitational action.
One first notices that in pure Einstein gravity, with Lagrangian density 
\beq
{\cal L} \; = \; - \, { 1 \over 16 \pi \, G} \, \sqrt{g} \, R \;\; ,
\eeq
the bare coupling $G$ can be completely reabsorbed by a suitable field redefinition
\beq
g_{\mu\nu} = \omega \, g_{\mu\nu}'
\label{eq:metric-scale}
\eeq
with $\omega$ a constant.
It follows that in a quantum formulation the renormalization
properties of $G$ have no physical meaning for this theory, 
at least until some other terms are added, to be discussed below.
The reason of course is that the term $\sqrt{g} R$ 
is homogeneous in $g_{\mu\nu}$, which is quite different from the Yang-Mills case. 
The situation changes though when one introduces a second dimensionful
quantity to compare with.
In the gravity case, this contribution is naturally
supplied by matter, or by a cosmological constant term proportional to $\lambda_0$,
\footnote{
In the following we will denote by $\lambda_0$ the (un-scaled) cosmological
constant, and by $\lambda$ the scaled one, so that $\lambda_0 \equiv
\lambda / 8 \pi G$.
In this work the symbol $\Lambda$ will be reserved for the ultraviolet cutoff.}
\beq
{\cal L} \; = \; 
- { 1 \over 16 \pi \, G} \, \sqrt{g} \, R \, + \, \lambda_0 \,
\sqrt{g} \; .
\label{eq:lagrange}
\eeq
Under a rescaling of the metric, as in Eq.~(\ref{eq:metric-scale}), one
obtains
\beq
{\cal L} = - { 1 \over 16 \pi \, G} \, \omega^{d/2-1} \, \sqrt{g'} \, R' 
\, + \, \lambda_0 \, \omega^{d/2} \, \sqrt{g'} \; ,
\label{eq:rescale}
\eeq
which is seen as being equivalent to a rescaling of the two bare couplings
\beq
G \rightarrow \omega^{-d/2+1} G \; , \;\;\;\; 
\lambda_0 \rightarrow \lambda_0 \, \omega^{d/2} 
\label{eq:rescale_g}
\eeq
while at the same time leaving the dimensionless 
combination $G^d \lambda_0^{d-2}$ unchanged.
Therefore only the latter quantity has physical meaning in pure
gravity, and it would seem physically meaningless here
to discuss separately the renormalization properties
of $G$ and $\lambda_0$.
In particular, one can always choose the scale $\omega = \lambda_0^{-2/d}$,
so as to adjust the volume term to have a unit coefficient.
Then one obtains 
\beq
{\cal L} = - { 1 \over 16 \pi \, G \, \lambda_0^{1-2/d} } \, \sqrt{g'} \, R' 
\, + \, \sqrt{g'} \; .
\label{eq:scaled}
\eeq
One concludes that the only coupling that matters for pure gravity in 
four dimensions is $G \sqrt{\lambda_0}$, so that, without any loss of 
generality, it would seem one can take $\lambda_0=1$ in units of some UV cutoff.

Nevertheless, a discussion of the field rescaling properties of the theory
is incomplete unless one also takes into account the effect of
the functional measure.
Following DeWitt \cite{dew67}, one defines an invariant norm for metric
deformations as
\beq
\Vert \delta g \Vert^2 \, = \, 
\int d^d x \; \delta g_{\mu \nu}(x) \;
G^{\mu \nu, \alpha \beta} \bigl [ g(x) \bigr ] \;
\delta g_{\alpha \beta}(x) \;\; ,
\label{eq:dw-def}
\eeq
with the supermetric $G$ given by the ultra-local 
(since it is defined at a single point $x$) expression
\beq
G^{\mu \nu, \alpha \beta} \bigl [ g(x) \bigr ] \, \equiv \, 
\half \, \sqrt{g(x)} \, \left [ \,
g^{\mu \alpha}(x) \, g^{\nu \beta}(x) +
g^{\mu \beta}(x) \, g^{\nu \alpha}(x) + \lambda \,
g^{\mu \nu}(x) \, g^{\alpha \beta}(x) \, \right ]
\label{eq:dw-super}
\eeq
and $\lambda$ a real parameter such that $\lambda \neq - 2 / d $.
The above supermetric then defines a suitable volume element
$\sqrt{\det G}$ in function space, and the functional 
measure over the $g_{\mu\nu}$'s takes on the form
\beq
\int [d \, g_{\mu\nu} ] \, \equiv \, \int \, \prod_x \, 
\Bigl [ \, \det G [ g(x) ] \, \Bigr ]^{1/2} \,
\prod_{\mu \geq \nu} d g_{\mu \nu} (x) \;\; .
\label{eq:dw-det0}
\eeq
The assumed locality of the supermetric 
$G^{\mu \nu, \alpha \beta} [ g(x) ] $ 
implies that its determinant is also a local function of $x$ only.
Up to an inessential multiplicative constant one finds
\beq
\int [d \, g_{\mu\nu} ] \, = \, 
\int \, \prod_x \, \bigl [ g(x) \bigr ]^{ (d-4)(d+1)/8 } \,
\prod_{\mu \ge \nu} \, d g_{\mu \nu} (x)
\; \mathrel{\mathop\rightarrow_{ d \rightarrow 4 }} \;
\int \, \prod_x \, \prod_{\mu \ge \nu} \, d g_{\mu \nu} (x) \; .
\label{eq:dw-dewitt}
\eeq
However it is not obvious that the above construction 
of the measure is unique;
an alternative derivation starts from a slightly different
supermetric, and leads to the scale-invariant functional measure
\beq
\int [d \, g_{\mu\nu} ] \,  = \, 
\int \, \prod_x \, \bigl [ g(x) \bigr ]^{ -d(d+1)/8 } \,
\prod_{\mu \ge \nu} \, d g_{\mu \nu} (x)
\; \mathrel{\mathop\rightarrow_{ d \rightarrow 4 }} \;
\int \, \prod_x \, \bigl [ g(x) \bigr ]^{ -5/2 } \,
\prod_{\mu \ge \nu} \, d g_{\mu \nu} (x) \; ,
\label{eq:misner}
\eeq
which was originally suggested in \cite{mis57}.
For a more complete discussion of the many delicate issues 
associated with the formulation of the covariant Feynman 
path integral approach to quantum gravity the reader 
is referred to \cite{book}. 
One can further show that if one introduces an $n$-component
scalar field $\phi(x)$ in the functional integral, it leads to
more changes in the gravitational measure.
For the functional measure over $\phi$ one writes
\beq
\int [ d \phi ] \, = \,
\int \prod_x \, 
\bigl [ \sqrt{g(x)} \bigr ]^{n/2} \, \prod_x \, d \phi(x) \;\; ,
\eeq
so that the first factor gives an additional contribution
to the gravitational measure.
These arguments lead one to conclude that the volume factor
in the measure should be written more generally as $g^{\sigma /2}$, 
and be included in a slightly more general form for the gravitational 
functional measure, as given before in Eq.~(\ref{eq:measure}).
In principle there is no clear a priori way of deciding between the
various choices for $\sigma$, and it may very well turn
out to be an irrelevant parameter.
The only constraint seems that the regularized gravitational
path integral should be well defined,
which would seem to rule out singular measures.
It is noteworthy though that the $g^{\sigma/2}$ volume term in the measure is 
completely local and contains no derivatives,
which therefore cannot affect the propagation
properties of gravitons.
But more importantly, for our purpose here it will be sufficient 
to note that under a rescaling of the metric 
the functional measure in Eq.~(\ref{eq:measure})
only picks up an irrelevant multiplicative constant.
The fact that the latter depends on the specific form of
the functional measure (i.e. on $\sigma$) is completely irrelevant;
such a constant drops out automatically when computing averages.
One reaches therefore the conclusion that the previous 
arguments remain largely unchanged: the functional integral for
pure gravity only depends on one dimensionless combination
of $G$ and $\lambda_0$; $\lambda_0$ can be set equal to unity
in units of th UV cutoff without any loss of generality \cite{hw84,lesh84}.

From a physical perspective, it might nevertheless seem more appropriate
to keep the dimensions of various parameters appearing in
the action unchanged.
This can be achieved by explicitly introducing an ultraviolet cutoff 
$\Lambda$, so that the Euclidean Einstein-Hilbert action with a
cosmological term is written in four dimensions as
\beq
I \; = \; \lambda_0 \, \Lambda^4 \int d^4 x \, \sqrt{g} \, - \, 
{ \Lambda^2 \over 16 \pi \, G } \int d^4 x \, \sqrt{g} \, R \; .
\label{eq:ehaction}
\eeq
In this expression $\lambda_0$ is the bare cosmological constant
and $G$ the bare Newton's constant, both now written in units 
of the explicit ultraviolet cutoff $\Lambda$.
Consequently, both of the above $G$ and $\lambda_0$ are now
dimensionless [a natural expectation is for the bare microscopic, 
dimensionless couplings not to be fine-tuned, and have magnitudes 
of order one, $\lambda_0 \sim G \sim O(1) $].
Now, again one can rescale the metric
\beq
g_{\mu\nu}' \; = \; \sqrt{\lambda_0} \; g_{\mu\nu}
\;\;\;\;\; 
{g'}^{\mu\nu} \; = \; { 1 \over \sqrt{\lambda_0} } \; g^{\mu\nu}
\label{eq:metric-scale-1}
\eeq
and thus obtain for the action
\beq
I \; = \; \Lambda^4 \int d^4 x \, \sqrt{g'} \, - \, 
{ \Lambda^2 \over 16 \pi \, G \, \sqrt{\lambda_0} } \, \int d^4 x \,
\sqrt{g'} \, R' \; .
\eeq
The latter, for a given cutoff $\Lambda$, only depends on 
$\lambda_0$ and $G$ through the dimensionless combination 
$G \sqrt{\lambda_0}$.
Next consider again the Euclidean Feynman path integral 
of Eq.~(\ref{eq:zcont}), here in four dimensions
\beq
Z \; = \; \int [ d \, g_{\mu\nu} ] \; \exp \left \{ 
- \int d^4 x \, \sqrt g \; \Bigl 
( \lambda_0 \, \Lambda^4 \, - \, { \Lambda^2 \over 16 \pi \, G } \, R \,
\Bigr ) \right \} \; .
\label{eq:zcont1}
\eeq
Because of the scaling properties of the functional measure
over metrics, and for a given cutoff $\Lambda$, 
$Z$ itself also depends, up to an irrelevant overall multiplicative constant,
on $\lambda_0$ and $G$ only through the dimensionless combination 
$G \sqrt{\lambda_0}$; only the latter can play a role in the 
subsequent physics.
One can then view a rescaling of the metric as simply a 
(largely inessential) redefinition of the ultraviolet cutoff 
$\Lambda$, $\Lambda \rightarrow \lambda_0^{-1/4} \Lambda$.
Furthermore, the existence of a non-trivial
ultraviolet fixed point for quantum gravity in four
dimensions is entirely controlled
by this dimensionless parameter only, both on the lattice \cite{hw84,lesh84}
and in the continuum \cite{eps}.

It is clear from the structure of the path integral 
that the cosmological term controls the overall scale in the
problem, while the curvature term provides the necessary derivative,
or true coupling, term.
But by a specific choice of overall scale one can set, without any
loss of generality $\lambda_0 = 1$
in Eq.~(\ref{eq:zcont1}), and measure from now on all quantities
in units of the UV cutoff $\Lambda$, which is the only
remnant of this overall scale.
\footnote{
We recall here that our considerations here are not dissimilar from the case of a 
self-interacting scalar field, where one might want to introduce three
couplings for the kinetic term, the mass term and the quartic
coupling term, respectively. A simple rescaling of the field
then reveals immediately that only two coupling ratios are in fact
physically relevant.}
In addition, one can choose for ease of notation a unit cutoff
$\Lambda=1$, 
and later restore, if one so desires, the correct dimensionality of 
couplings and operators by
suitably re-introducing appropriate powers of the UV cutoff $\Lambda$.
Indeed this is a common, if not universal, procedure in lattice field theory 
and lattice gauge theory, where all quantities are measured
in terms of a unit lattice spacing $a$.
Since the total volume of space-time 
can hardly be considered a physical observable, quantum
averages are in fact computed by dividing out by the total
volume.
Thus, for example, for the quantum expectation value of the Ricci scalar one writes
\beq
{\cal R } \; \equiv \; 
{ \langle \, 
\int d^4 x \, {\textstyle {\sqrt{g(x)}} \displaystyle} \; R(x) 
\, \rangle
\over 
\langle \, \int d^4 x \, {\textstyle {\sqrt{g(x)}} \displaystyle} \,
\rangle } \; .
\label{eq:average}
\eeq

The discussion so far has focused on pure gravity without matter
fields.
The addition of matter prompts one to do some further rescalings.
Let us consider here for simplicity, and as an illustration,
a single component scalar field, with action given by
\beq
I_S \; = \; \half \, \int d^4 x \sqrt{g} \, \left \{ 
g^{\mu\nu} \, \partial_\mu \, \phi \, \partial_\nu \, \phi \, + \, m_0^2 \, \phi^2
\, + \, R \, \phi^2  \right \}
\eeq
and functional measure for $\phi$ 
\beq
\int d \mu [\phi] \; = \; \int \prod_x \,
\left [ g(x) \right ]^{1/4} \, d \phi (x) \; .
\eeq
Then, if the the metric rescaling of Eq.~(\ref{eq:metric-scale-1})
is followed by a field rescaling
\beq
\phi ' (x) \; = \; { 1 \over \lambda_0^{1/4} } \, \phi (x) \; ,
\eeq
one sees that the only surviving change is a rescaling of the bare mass,
$m_0 \rightarrow m_0 \, \lambda_0^{1/4} $. 
In addition, the scalar field functional measure acquires an
irrelevant multiplicative factor, which as stated before
cannot affect quantum averages.

Let us add here one further comment.
Pure gravity corresponds to a massless graviton, a property that
is presumably preserved to all orders in perturbation theory,
if diffeomorphism invariance is maintained.
Nevertheless it is easy to see that formally the cosmological term,
at least in the weak field limit, acts as a mass-like term.
In the weak field expansion around flat space one has for the cosmological contribution 
in Eq.~(\ref{eq:lagrange})
\beq
\sqrt g = 1 + \half \, h_\mu^{\;\;\mu}
+ \eigth \, h_\mu^{\;\;\mu} h_\nu^{\;\;\nu}
- \quarter \, h_{\mu\nu} h^{\mu\nu} + O(h^3) \; ,
\label{eq:h-cosm}
\eeq
after setting $ g_{\mu\nu} = \eta_{\mu\nu} + h_{\mu\nu} $
with $\eta_{\mu\nu}$ the flat metric.
The first contribution $O(h)$ shifts the vacuum solution
to de Sitter space, while the next terms $O(h^2)$ provide
a mass-like quadratic contributions.
It is tempting therefore to still regard, in some ways, this last
term as analogous to some sort of mass term, with 
nevertheless the rather important property that 
it does not lead to an explicit breaking of general covariance.
One would then expect that such a mass-like term could
provide naturally, in a general renormalization framework,
a suitable candidate for a renormalization 
group invariant quantity, in analogy to the dynamically generated 
mass parameter in non-Abelian gauge theories, which also
generates a dynamical mass, without violating local gauge invariance.
How this might come about will be expanded on further below.

\vskip 30pt

\section{Gauge Dependence in the Renormalization of the Cosmological Constant}

\label{sec:pert}

\vskip 10pt

Perturbation theory generally serves a very useful purpose, since it allows
one to systematically track the gauge dependence of various 
renormalization effects, and determine what is physical
and what is a spurious gauge artifact.
Unfortunately Einstein gravity is not perturbatively renormalizable
in four dimensions, so that easy route is not available. 
Nevertheless, if one goes down in dimensions it is possible
to rescue in part the perturbative treatment, and therefore
address some of the key issues raised earlier.
One does not, of course, expect the answers to be quantitatively correct;
nevertheless it will become clear below that the issue of gauge
invariance does comes up, and is eventually successfully resolved.
Let us emphasize here that one key aspect of the perturbative 
treatment via the background field method is that diffeomorphism
invariance is preserved throughout the calculation,
as in the case of non-Abelian gauge theories.

In two dimensions the gravitational coupling is dimensionless, 
$G\sim \Lambda^{2-d}$ and the theory appears perturbatively renormalizable.
In spite of the fact that the gravitational action reduces to a topological
invariant in two dimensions, it is meaningful to try to construct, 
in analogy to Wilson's original suggestion for scalar field theories,
the theory perturbatively as a double series in $\epsilon=d-2$ and $G$
\cite{wei79,gas78}.
The $2+\epsilon$ expansion for pure gravity then proceeds as follows \cite{eps}.
First the gravitational part of the action
\beq
{\cal L} = - { \mu^\epsilon \over 16 \pi \, G} \, \sqrt{g} \, R \;\; ,
\label{eq:l-pure}
\eeq
with $G$ dimensionless and $\mu$ an arbitrary momentum scale, 
is expanded in the fields by setting
\beq
g_{\mu\nu} \, \rightarrow \, \bar g_{\mu\nu} = g_{\mu\nu} \, + \, h_{\mu\nu}
\eeq
where $g_{\mu\nu}$ is the classical background field and $h_{\mu\nu}$
the small quantum fluctuation.
The quantity ${\cal L}$ in Eq.~(\ref{eq:l-pure}) is naturally identified with
the bare Lagrangian, and the scale $\mu$ with a microscopic ultraviolet
cutoff $\Lambda$; the latter would be identified with the inverse 
lattice spacing in a lattice formulation.
To make perturbation theory convergent requires a gauge fixing term,
chosen by the authors of \cite{eps} in the form of a generalized background 
harmonic gauge condition,
\beq
{\cal L}_{gf} \, = \, \half \, \alpha \, 
\sqrt{ g} \;  g_{\nu\rho}
\left ( 
\nabla_\mu h^{\mu\nu} - \half \, \beta \, g^{\mu\nu} \nabla_\mu h 
\right )
\left ( 
\nabla_\lambda h^{\lambda\rho} - \half \, \beta \, g^{\lambda\rho} \nabla_\lambda h 
\right )
\label{eq:l-gauge}
\eeq
with $ h^{\mu\nu} = g^{\mu\alpha} g^{\nu\beta} h_{\alpha\beta} $,
$h = g^{\mu\nu} h_{\mu\nu}$ and $ \nabla_\mu $ the 
covariant derivative with respect to the background metric
$g_{\mu\nu}$.
The gauge fixing term then requires the introduction of a 
Faddeev-Popov ghost contribution ${\cal L}_{ghost}$ 
containing the ghost field $\psi_\mu$, so that the total
Lagrangian becomes a sum of three terms,
${\cal L}+{\cal L}_{gf}+{\cal L}_{ghost}$.
In a flat background, $ g^{\mu\nu} = \delta^{\mu\nu}$,
one obtains, from the quadratic part of the Lagrangian of
Eqs.~(\ref{eq:l-pure}) and (\ref{eq:l-gauge}),
a rather complicated expression for the graviton propagator \cite{eps}
\bea
\langle \, h_{\mu\nu} (k) h_{\alpha\beta} (-k) \, \rangle \,& = &
{1 \over k^2 } \, ( \delta_{\mu\alpha} \delta_{\nu\beta} +
\delta_{\mu\beta} \delta_{\nu\alpha} )
\, - \, { 2 \over d-2 } \, { 1 \over k^2 } \, \delta_{\mu\nu} \delta_{\alpha\beta}
\nonumber \\
&& - \left ( 1 \! - \! {1 \over \alpha } \right ) \, {1 \over k^4 } \,
( \delta_{\mu\alpha} k_\nu k_\beta + 
\delta_{\nu\alpha} k_\mu k_\beta  +
\delta_{\mu\beta} k_\nu k_\alpha  +
\delta_{\nu\beta} k_\mu k_\alpha )
\nonumber \\
&& + \, { 1 \over d-2 } \,
{ 4 (\beta-1) \over \beta - 2 } \, { 1 \over k^4 } \,
( \delta_{\mu\nu} k_\alpha k_\beta + \delta_{\alpha\beta} k_\mu k_\nu )
\nonumber \\
&& + \; { 4 \, (1-\beta) \over (\beta-2)^2 } \, \left [ 
2 - { 3 - \beta \over \alpha } - { 2 \, (1-\beta) \over d-2 } \right ]
\, {1 \over k^6 } \, k_\nu k_\nu k_\alpha k_\beta  \;\; .
\label{eq:prop-2d}
\eea
Normally it would be convenient to choose a gauge $\alpha=\beta=1$,
in which case only the first two terms for the graviton
propagator survive \cite{gas78}.
But it is in fact rather advantageous
to leave the two gauge parameters unspecified,
so that later the detailed gauge dependence of the result can be checked.
In particular, the gauge parameter $\beta$ is related
to the gauge freedom associated with a rescaling 
the metric $g_{\mu\nu}$, as described in the previous section.
For the one loop divergences associated with the $\sqrt{g}$ term
they obtain
\beq
\lambda_0 \rightarrow \lambda_0 \left [ \, 
1 - \left ( 
{ a_1 \over \epsilon } + 
{ a_2 \over \epsilon^2 } \right ) G \, \right ] \; ,
\label{eq:a-co}
\eeq
with coefficients
\bea
a_1 & = & - { 8 \over \alpha } + 8 \, { (\beta-1)^2 \over (\beta-2)^2 } 
+ 4 \, {( \beta -1 ) (\beta -3) \over \alpha \, (\beta-2)^2 }
\nonumber \\
a_2 & = & 8 \, { (\beta-1)^2 \over (\beta-2)^2 } \; .
\label{eq:a-coeff}
\eea
On the other hand, for the one-loop divergences associated with 
the $\sqrt{g} R $ term one finds
\beq
{ \mu^\epsilon \over 16 \pi \, G}
\rightarrow
{ \mu^\epsilon \over 16 \pi \, G} \left ( 1 - { b \over \epsilon } \, G \right )
\eeq
with coefficient $b$ given by \cite{gas78}
\beq
b = { 2 \over 3 } \cdot 19 + { 4 (\beta-1)^2 \over (\beta-2)^2 } \; .
\label{eq:b-coeff}
\eeq
Thus the one-loop radiative corrections modify the total Lagrangian to
\beq
{\cal L} \rightarrow
- { \mu^\epsilon \over 16 \pi \, G} \left ( 1 - { b \over \epsilon } G \right )
\sqrt{g} \, R 
+ \lambda_0 \left [ \, 1 - \left (  
{ a_1 \over \epsilon } + { a_2 \over \epsilon^2 } \right ) G \, \right
] \sqrt{g} \; .
\eeq
Next one can make use of the freedom to rescale the metric, by setting
\beq
\left [ \, 1 - \left (  
{ a_1 \over \epsilon } + { a_2 \over \epsilon^2 } \right ) G \, \right ] \sqrt{g}
\, = \, \sqrt{g'} \; ,
\label{eq:scaled-cosm}
\eeq
which restores the original unit coefficient for the cosmological constant term.
The rescaling is achieved by the field redefinition
\beq
g_{\mu\nu} \, = \,
\left [ 1 - \left (  
{ a_1 \over \epsilon } + { a_2 \over \epsilon^2 } \right ) G \right ]^{-2/d} 
\, g'_{\mu\nu} \; .
\eeq
By this procedure the cosmological term is brought back into its standard form
$\lambda_0 \sqrt{g'}$, and one obtains for the
complete Lagrangian to first order in $G$
\beq
{\cal L} \rightarrow
- { \mu^\epsilon \over 16 \pi \, G} 
\left [ 1 - { 1 \over \epsilon } ( b - \half a_2 ) G \right ]
\sqrt{g'} \, R' + \lambda_0 \sqrt{g'} \; ,
\eeq
where only terms singular in $\epsilon$ have been retained.
From this last result one can finally read off the renormalization of 
Newton's constant
\beq
{ 1 \over G} \rightarrow { 1 \over G }
\left [ \, 1 - { 1 \over \epsilon } ( b - \half a_2 ) \, G \, \right ]
\; .
\label{eq:g-ren}
\eeq
From Eqs.~(\ref{eq:a-coeff}) and (\ref{eq:b-coeff}) one notices
that the $a_2$ contribution cancels out the gauge-dependent
part of $b$, giving for the remaining contribution
$ b - \half a_2 = \twoth \cdot 19 $.
Therefore the gauge dependence has, as one would have hoped for on physical
grounds, entirely disappeared from the final answer.
The reason for this miraculous cancellation is of course
general covariance.
But the main point we wish to make here is that the results of
covariant perturbation theory are, as expected, entirely consistent
with the scaling argument given in the previous section:
only the renormalization of $G$ has physical meaning.
Let us dwell further on this aspect. 

In the presence of an explicit renormalization scale parameter
$\mu$ the Callan-Symanzik $\beta$-function for pure gravity is obtained by
requiring the independence of the effective coupling $G$
from the original renormalization scale $\mu$.
One obtains to one loop order
\beq
\mu { \partial \over \partial \mu } \, G ( \mu ) \, \equiv \,
\beta (G) = (d-2) \, G \, - \, \beta_0 \, G^2 \, + \,
O( G^3, (d-2) G^2 )
\label{eq:beta-oneloop}
\eeq 
with here $\beta_0 = \twoth \cdot 19 $ in the absence of matter.
From the procedure outlined above it is clear that $G$ is
the only coupling that is scale-dependent in pure gravity.
Depending on whether one is on the right ($G>G_c$) or on the left 
($G<G_c$) of the non-trivial ultraviolet fixed point at
\beq
G_c = { d - 2 \over \beta_0 } + O((d-2)^2 )
\eeq
the coupling will either flow to increasingly larger values of $G$,
or flow towards the Gaussian fixed point at $G=0$, respectively.
In the following we will refer to the two phases as the strong
and weak coupling phase, respectively.
Perturbatively one only has control on the small $G$ regime. 

The running of $G$ as a function of a sliding momentum scale $\mu=k$
in pure gravity is obtained from integrating 
Eq.~(\ref{eq:beta-oneloop}), giving
\beq
G( k ) \; \simeq \; G_c \, \left [ 
\, 1 \, \pm \, c_0 \, \left ( {m^2 \over k^2 } \right )^{(d-2)/2}
\, + \, \dots \right ]
\label{eq:grun_k} 
\eeq
with $c_0$ a positive constant, and $m = \xi^{-1}$ a mass scale
that arises as an integration constant of the renormalization
group equations.
The $k^2$-dependent contribution on the r.h.s of 
Eq.~(\ref{eq:grun_k}) is the quantum correction, which at
least within a perturbative framework is assumed to be small.
The choice of $+$ or $-$ sign is determined from whether one is
to the left (-), or to right (+) of $G_c$, in which case
the effective $G(k)$ decreases or, respectively, increases as one flows away
from the ultraviolet fixed point towards lower momenta or larger distances.
Physically the two solutions represent a screening ($G<G_c$) and an 
anti-screening ($G>G_c$) situation.
While in the above continuum perturbative calculation both phases,
and therefore both signs, seem acceptable, the Euclidean 
and Lorentzian lattice
results on the other hand rule out the weak coupling
phase as pathological, in the sense that there the lattice collapses into a
two-dimensional degenerate object \cite{hw84,book}.

The $k^2$-dependent quantum correction in Eq.~(\ref{eq:grun_k})
involves a new physical, renormalization group invariant scale $\xi=1/m$
which cannot be fixed perturbatively, and whose size then determines the 
distance scale relevant for quantum effects.
In terms of the bare coupling $G(\Lambda)$, it is given by
\beq
\xi^{-1} (G) \, \equiv \, m \, = \, \Lambda \cdot A_m \, 
\exp \left ( { - \int^{G(\Lambda)} \, {d G' \over \beta (G') } } \; ,
\right )
\label{eq:m-cont}
\eeq
with $A_m$ a constant.
Note the rather remarkable fact that one scale has disappeared ($\lambda_0$),
and a new one has appeared dynamically ($\xi$).
The above expression is obtained by integrating the RG equation
$ \mu \, { \partial \over \partial \mu } \, G = \beta (G)$,
and then choosing the arbitrary momentum scale $\mu \rightarrow \Lambda$.
Conversely, $\xi^{-1} = m$ is an RG invariant and one has 
\beq
\Lambda \, { d \over d \Lambda } \, m ( \Lambda, G(\Lambda) )
\; = \; \mu \, { d \over d \mu } \, m ( \mu, G (\mu ) ) \; = \; 0 \; .
\label{eq:m-invariance}
\eeq
The running of $G(\mu)$ in accordance with the renormalization
group equation of Eq.~(\ref{eq:beta-oneloop}) ensures that the l.h.s. 
is indeed a renormalization group invariant.
One knows that the constant $A_m$ on the r.h.s. of Eq.~(\ref{eq:m-cont})
cannot be determined perturbatively,
it needs to be computed by nonperturbative (lattice) methods,
for example by evaluating invariant correlations at fixed
geodesic distances; 
it is related to the constant $c_0$ in Eq.~(\ref{eq:grun_k})
by $c_0 = 1 / ( A_m^{1/\nu} G_c ) $.
In the vicinity of the ultraviolet fixed point at $G_c$, for
which $\beta(G_c)=0$, one can write
\beq
\beta (G) \, \equiv \, 
\mu \, { \partial \over \partial \, \mu } \, G( \mu )
\; \mathrel{\mathop\sim_{ G \rightarrow G_c }} \;
\beta ' (G_c) \; (G - G_c) \, + \, \dots \;\; ,
\label{eq:beta-g}
\eeq
which by integration gives
\beq
\xi^{-1} (G)  \; \propto \;
\Lambda \,  | \,  ( G - G_c ) / G_c \, |^{\nu } \; ,
\label{eq:xi_gc}
\eeq
with correlation length exponent $\nu = - 1 / \beta'(G_c) $;
to lowest order perturbation theory $\nu = 1 / (d-2) + \dots $.
Note that the magnitude of $\xi$ is not determined by the
magnitude of $G$.
Instead, it is determined by the distance of the bare $G$ from
the UV fixed point value $G_c$, and as such it can 
be {\it arbitrarily} large.

More recently the one-loop perturbative calculation described
above were laboriously extended to two loops \cite{ak97}.
One important result that stays true to two loops is the
fact that the only meaningful, gauge-independent renormalization 
is the one of $G$, which is not surprising in view of
the general arguments given previously.
One can then compute the roots $\beta (G_c) \, = \, 0 $ and obtain the location
of the ultraviolet fixed point, and from it
the universal exponent $\nu = -1 /\beta'(G_c)$.
One finds for the scaling exponent $\nu$ in the presence of $c$ scalar matter fields
\beq
\nu^{-1} \; = \; 
(d-2) \, + \, {15 \over 25 - c } \, (d-2)^2 \, + \, \dots \; .
\label{eq:nueps}
\eeq
In four dimensions this gives for pure gravity without
matter ($c=0$) to lowest order
$\nu = 1/2$, and $\nu =5/22 \approx 0.23 $ at the next order.
Numerical simulations for the lattice theory of gravity in
four dimensions give on the other hand 
$\nu=1/3 $ and $c_0 \approx 8.0 $ \cite{ham00}.

In closing we mention that, so far, the discussion of quantum gravity
has focused mainly on the perturbative scenario, 
where the gravitational coupling $G$ is assumed
to be weak, so that the weak field expansion can be pursued with some
degree of reliability.
Then at every order in the loop expansion the problem reduces 
to the evaluation of an increasingly complicated sequence of Gaussian
integrals over some small quantum fluctuation in the fields.
Nevertheless a bit of thought reveals that to all orders
in the weak field expansion there is really no difference of
substance between the Lorentzian (or pseudo-Riemannian) and the
Euclidean (or Riemannian) formulation.
The structure of the divergences would have been identical,
and the renormalization group properties of the coupling
the same (up to the trivial replacement of say the Minkowski momentum
$q^2$ by its Euclidean expression $q^2 = q_0^2 + {\bf q}^2 $ etc.).
Thus, up to this point, no significant difference has appeared between the
Euclidean and the Lorentzian treatment.
We should also add that most of the above conclusions remain largely
unchanged when higher derivative terms are included \cite{fra81,jul78,avr85}.

To summarize the results so far, we have shown that the path integral
for pure quantum gravity depends only on one dimensionless
combination of couplings, $G \sqrt{\lambda_0}$ in $d=4$, and that
the bare $\lambda_0$ can be entirely scaled out of the path integral, 
and out of the physics.
It is also clear that the only renormalization (and beta function)
that is gauge-independent and physically meaningful is the one for
Newton's constant $G$.
Finally, we have emphasized the fact that the very same, manifestly
covariant, renormalization group treatment clearly shows the appearance 
of a new dynamically generated scale $\xi$ [Eq.~(\ref{eq:m-cont})].

\vskip 30pt

\section{Lattice Functional Integral and Role of the Volume Term}

\label{sec:latt}

\vskip 10pt

One might view some of the discussion of the previous sections as
rather formal. 
The Feynman path integral for quantum gravitation,
Eq.~(\ref{eq:zcont}), is formally defined in the continuum and
involves rather delicate expressions such as $\prod_x $ in the measure.
Perturbation theory in the continuum is then done by performing
Gaussian integrals for small metric perturbations,
using dimensional regularization to manipulate the resulting divergent integrals.
It seems useful therefore to revisit here the same kind 
of issues, as they arise in the context of the lattice theory.
In the Regge-Wheeler formulation of lattice gravity \cite{reg61,whe64} 
the infinite number of degrees of freedom of the continuum
gravitational field is restricted by considering piecewise-linear
Riemannian spaces described by a finite number of variables, 
the geodesic distances between neighboring points.
It is known to be the only lattice formulation of gravity containing
transverse-traceless modes \cite{rowi81} and a local continuous
lattice diffeomorphism invariance \cite{rowi81,gauge}.
An Euclidean path integral formulation can then be built based on
a curvature action, supplemented by a cosmological term for convergence, 
and possibly higher derivative contributions \cite{hw84,lesh84}.
Following Regge, one writes for the Euclidean lattice action for pure gravity
\beq
I_{R}  \; = \; - \; k \, \sum_{\rm hinges \; h}
\, \delta_h (l^2) \, A_h  (l^2)\;\; ,
\label{eq:regge}
\eeq
with bare coupling constant $k^{-1} = 8 \pi \, G $. 
In four dimensions the sum over hinges $h$ is equivalent to a sum
over all lattice triangles;
geometrically, the above action contains a sum of elementary 
loop contributions, since it contains as its primary ingredient the deficit
angle $\delta_h $ associated with an elementary parallel transport loop
around the hinge $h$.
The deficit angle $\delta_h$ is related to the local scalar curvature by
\beq
R (h) \; = \; 2 \; { \delta_h \over A_C (h) }
\label{eq:r-hinge}
\eeq
where $A_C (h) $ is the area associated with an elementary
parallel transport loop around the hinge (triangle) $h$,
defined by joining the vertices of an elementary polyhedron $C$
located in the dual lattice.
In view of the following discussion one should note
that, as in the continuum, the local lattice curvature has dimensions
of length to the power minus two.
The continuum curvature density $\sqrt{g} \, R$ is then obtained
by multiplication with the volume element $V_h $ associated with
a hinge $h$, with the lattice Riemann tensor at a hinge $h$ given by
\beq
R_{\mu\nu\lambda\sigma} (h) \; = \; {\delta_h  \over A_C (h) } 
\, U_{\mu\nu} (h) \, U_{\lambda\sigma} (h) \; .
\label{eq:riem-hinge}
\eeq
Here $U_{\mu\nu} (h) $ is unit bivector defined for a single hinge $h$ 
\beq
U_{\mu\nu} (h) \; = \; { 1 \over 2 A_h } \;
\epsilon_{\mu\nu\alpha\beta} \, l_1^{\alpha} \, l_2^{\beta} \; ,
\label{eq:bivector}
\eeq
with $l_1 (h)$ and $l_2 (h)$ two independent edge vectors
associated with the hinge (triangle) $h$, and $A_h$ the area 
of the hinge itself (a triangle in four dimensions).
Again it is customary, as in lattice gauge theories,
to set the lattice ultraviolet cutoff equal to one
(i.e. measure all length scales in units of a fundamental
lattice cutoff $a$ or $\Lambda$; as an example, on a hypercubic lattice in
$d$ dimensions the two are simply related by $ \Lambda = \pi / a $).
Next consider the cosmological constant term, which in the continuum theory
takes the form $ \lambda_0 \int d^d x \sqrt g $.
On the lattice it involves the total volume of the simplicial complex
\beq
V \;  = \; \sum_{\rm simplices \; s} V_s (l^2) \; .
\label{eq:totaltvol}
\eeq
In four dimensions the sum here is over all lattice four-simplices, 
the 4d analogs of tetrahedra.
Thus one may regard the local volume element in d dimensions,
$ \sqrt g \, d^d x $, as
being represented by $ V_s$, centered on the simplex $s$.
The curvature and cosmological constant terms then lead to the combined action
\beq
I_{\rm latt} \; = \; \lambda_0 \, \sum_{\rm simplices \; s} \, V_s (l^2)
\, - \, k \, \sum_{\rm hinges \; h} \,  \delta_h (l^2) \, A_h (l^2) \; .
\label{eq:latac}
\eeq
The action then only couples edges which belong either to
the same simplex or to a set of neighboring simplices, and is
therefore local, as the continuum action. 

A lattice regularized version of the Euclidean Feynman path integral
is then given by
\beq
Z_{\rm latt} \; = \;  \int [ d \, l^2 ] \, \exp \left ( - I_{\rm latt} (l^2)
\right ) \;\; ,
\label{eq:zlatt}  
\eeq
where $ [ d \, l^2 ] $ is an appropriate functional integration
measure over squared edge lengths $ l_{ij}^2 $.
For concreteness, the local functional measure will be here of the form 
\beq
\int [ d \, l^2 ] \; \equiv \;
\int_0^\infty \; 
\prod_{ ij } \, dl_{ij}^2 \; 
\prod_s \; \left [ V_s (l^2) \right ]^{\sigma} \, \Theta [l_{ij}^2]
\;\; .
\label{eq:lattmeas}
\eeq
The last expression represents a rather non-trivial 
quantity, both in view of the complexity of the formula for the volume of a simplex,
and because of the generalized triangle inequality constraints 
implicit in $[d\,l^2]$, given that the function $ \Theta [l_{ij}^2] $ 
here represents a theta-function type constraint on the edge lengths.
The latter is inserted in order to ensure that the triangle
inequalities and their higher dimensional analogs are satisfied \cite{hw84,lesh84}.
The measure in Eq.~(\ref{eq:lattmeas}) should then be considered the 
lattice analogue of the gravitational DeWitt measure of Eq.~(\ref{eq:measure}).
Like the continuum functional measure, it is local
to the extent that each edge length
appears only in the expression for the volume of those simplices
which explicitly contain it.
Then the lattice partition function $Z_{latt}$ should in turn be
regarded as a discretized, and properly regularized, form of 
the continuum Euclidean Feynman path integral given in Eq.~(\ref{eq:zcont}).

As in the continuum, the curvature contribution to the lattice action
[Eq.~(\ref{eq:regge})] contains the proper kinetic (derivative) term, 
which then leads to a set of suitable propagating degrees of freedom,
the lattice transverse-traceless modes \cite{rowi81}.
Such a term provides the necessary coupling between neighboring lattice
metrics, nevertheless the interaction still remains local.
Moreover, due to the presence of the triangle area term $A_h$, the
curvature term in the action scales like a length squared: if all the
edge lengths are rescaled by a common factor $\omega$,
\beq
 l_i \; \rightarrow \; \omega \, l_i  \;\; ,
\label{eq:rescale_latt}
\eeq
then the curvature part of the action is simply rescaled 
by an overall factor of $\omega^2 $.
The latter can then be reabsorbed into a rescaling of the
coupling $G$, just as in the continuum [see Eq.~(\ref{eq:rescale})].
On the other hand, the cosmological term is just the
total four-volume of space-time.
As such, it does not contain any derivatives (or finite differences) 
of the metric and is completely local; 
it does not contribute to the propagation of gravitational degrees of
freedom and is thus more akin to a mass term (as is already clear
from the weak field expansion of $\int \sqrt{g}$ in the continuum).
This volume term scales like a length to the fourth power: if all the
edge lengths are rescaled by a common factor $\omega$,
$ l_i \rightarrow \omega \, l_i $, then the volume term
is simply rescaled by an overall factor of
$\omega^4 $.
Again, this effect can be entirely reabsorbed into a rescaling of 
the bare cosmological constant $\lambda_0$, as
in the continuum [see Eq.~(\ref{eq:rescale})].
\footnote{
Note that convergence of the Euclidean lattice functional integral
nevertheless requires a positive bare cosmological constant, $\lambda_0 > 0$.}

We note now that, as in the continuum case, the above considerations 
regarding the scaling properties
of the lattice gravitational action are not spoiled by
the functional lattice measure in Eq.~(\ref{eq:lattmeas}).
As one can see by inspection. when all the edge lengths
are rescaled by a common factor, the contribution 
from the functional measure is simply multiplied by a constant factor involving
$\omega$ to some power (which will depend on the overall number of
lattice points and on the choice measure parameter $\sigma$);
such a factor then drops out when evaluating expectation values.
More importantly, the overall length scale in the problem is
controlled by the parameter $\lambda_0$;
changing the value of $\lambda_0$ simply, and trivially, changes
this overall scale, without affecting in any way the 
underlying physics:
any change in $\lambda_0$ simply reflects itself in a change
in the average fundamental lattice spacing (or average local volume).
As such, this change is {\it physically irrelevant}.
Indeed, and in accordance with the methods of quantum field theory
and statistical field theory,
one would like to discuss renormalization group 
properties of the theory in a box of fixed total volume and fixed UV cutoff.
Allowing a change in the overall volume of the box, or
changing, equivalently, the value of the UV cutoff or lattice spacing, 
only hopelessly (and unnecessarily) confuses the whole renormalization issue. 
Of course, in a traditional renormalization group approach to field
theory, the overall four-volume is always kept fixed while the scale
(or $q^2$) dependence of the action and couplings are investigated.

It seems therefore again rather meaningless to allow the coupling $\lambda_0$ to run;
the overall space-time volume is intended to stay fixed within the RG
procedure, and not to be rescaled as well under a renormalization group transformation.
Indeed, in the spirit of Wilson, a renormalization group transformation
allows a description of the original physical system in terms
of a new coarse grained Hamiltonian, whose new operators are
interpreted as describing averages of the original system on a
finer scale - but of course still within the same very large physical
volume.
The only scale change in this procedure is from the coarse-scale averaging,
renormalization or block-spinning itself, to use here three roughly
equivalent terms.
The new effective Hamiltonian is then still supposed to
describe the original physical system, but does so more
economically in terms of a reduced set of effective degrees of freedom.
Consequently one can take the lattice coupling $\lambda_0 =1$ without
{\it any} loss of generality,  since different values of
$\lambda_0$ just correspond to a trivial rescaling of the overall
four-volume.
Alternatively, one could even choose an ensemble for
which the probability distribution in the total four-volume $V$ is
\beq
{\cal P} (V) \; \propto \; \delta ( V - V_0 ) \; ,
\eeq
in analogy with the microcanonical ensemble of statistical mechanics.
We conclude that the results from the lattice theory of gravity
{\it completely} mirror, and underpin, the discussion done for the
continuum theory in Sec.~(\ref{sec:cont}).
The lattice theory is shown to depend, in any dimension, on one coupling
only, the dimensionless combination of $G$ and $\lambda_0$;
in four dimensions this quantity is given by $G \sqrt{\lambda_0}$.
We have also given evidence that, without any loss of generality, one can 
take in the lattice regularized theory $\lambda_0=1$ in units of 
the UV cutoff, so that the theory depends simply on {\it one} coupling $G$ only.

On the lattice one finds that the running of $G$ is very similar
in structure to what is obtained in the $2+\epsilon$ expansion, even
though the procedure followed in obtaining such a result
is completely different:
on the lattice the scale dependence of $G$ is extracted from the
cutoff ($\Lambda$) dependence of the bare coupling $G(\Lambda$)
at fixed physical correlation length $\xi$ \cite{ham00}.
One finds, in the language of Eq.~(\ref{eq:grun_k}),
\beq
G( k ) \; \simeq \; G_c \, \left [ 
\, 1 \, + \, c_0 \, \left ( { \xi^2 \, k^2 } \right )^{- 1 / 2 \nu }
\, + \, \dots \right ]
\label{eq:grun_latt} 
\eeq
with $c_0$ a positive constant, $ \xi $ again the length scale
that arises as an integration constant of the renormalization
group equations, and exponent $\nu = 1/3 $.
Here the $k^2$-dependent correction can be compared directly
to the result given previously in Eq.~(\ref{eq:grun_k}).
Note that on the lattice only the $+$-sign is realized,
which corresponds to gravitational anti-screening.
This is because the Euclidean lattice results show
that the weak coupling phase ($G>G_c$) is pathological, 
corresponding not to a four-dimensional lattice but,
instead, to a two-dimensional degenerate object \cite{hw84,book}.

Here too the $k^2$-dependent quantum correction in Eq.~(\ref{eq:grun_latt})
involves a new nonperturbative renormalization group invariant scale
$\xi$, which determines the overall scale of quantum effects.
In terms of the bare coupling $G(\Lambda)$, it is given,
in the vicinity of fixed point $k_c$, by
\beq
\xi^{-1} (k) \; \; \mathrel{\mathop\sim_{ k \, \rightarrow \, k_c  }} \;
A_m \, \Lambda \,  | \,  k_c - k  \, |^{\nu } \; ,
\label{eq:xi_gc_latt}
\eeq
with correlation length exponent $\nu = 1/3 $.
A determination of the constant of proportionality 
$A_m$ in Eq.~(\ref{eq:xi_gc_latt})
then fixes the coefficient $c_0$ in the running of $G$
Eq.~(\ref{eq:grun_latt}), 
since the two coefficients are related by $c_0 = 1/ (k_c A_m^{1/\nu} )$.
By large scale numerical
simulations for the lattice theory of gravity in four dimensions one
finds explicitly for the critical point, the critical exponent and the 
amplitude, respectively,
$k_c = 0.06386(9)$, $\nu = 0.336(5)$ and
$c_0 = 8.02(55) $ \cite{ham00}.

\vskip 30pt

\section{The Gauge Theory Analogy}

\label{sec:gauge}

\vskip 10pt

$QED$ and $QCD$ provide two invaluable illustrative cases where the running
of the gauge coupling with energy is not only theoretically well
understood, but also verified experimentally.
As in $QED$, in $QCD$ (and related Yang-Mills theories) radiative
corrections are known to alter 
significantly the behavior of the static potential at short distances.
Changes in the potential are best expressed in terms 
of the running strong coupling constant $\alpha_S (\mu) $, whose scale
dependence is determined by the celebrated beta function of $SU(3)$ $QCD$
with $n_f$ light fermion flavors
\beq
\mu \, { \partial \, \alpha_S \over \partial \, \mu } \; = \; 
2 \, \beta ( \alpha_S ) \; = \; 
- \, { \beta_0 \over 2 \, \pi } \, \alpha_S^2 
\, - \, { \beta_1 \over 4 \, \pi^2 } \, \alpha_S^3 
\, - \, { \beta_2 \over 64 \, \pi^3 } \, \alpha_S^4 \, - \, \dots \; .
\label{eq:beta_qcd}
\eeq
with coefficients
$\beta_0 = 11 - {2 \over 3} n_f $, $\beta_1 = 51 - {19 \over 3} n_f $, 
and $\beta_2 = 2857 - {5033 \over 9} n_f + {325 \over 27} n_f^2 $, and
$\alpha_S \equiv g^2 / 4 \pi $.
The solution of the renormalization group equation  
Eq.~(\ref{eq:beta_qcd}) then gives for the running of $\alpha_S (\mu )$
\beq
\alpha_S (\mu ) \; = \; 
{ 4 \, \pi \over \beta_0 \ln { \mu^2 / \Lambda_{\overline{MS}}^2  } }
\, \left [ 1 \, - \, { 2 \beta_1 \over \beta_0^2 } \,
{ \ln \, [ \ln { \mu^2 / \Lambda_{\overline{MS}}^2 ] }
\over \ln { \mu^2 / \Lambda_{\overline{MS}}^2  } }
\, + \, \dots \right ] \; .
\label{eq:alpha_qcd}
\eeq
The nonperturbative scale $ \Lambda_{\overline{MS}} $ appears as an
integration constant of the renormalization group equations, and
is therefore - by construction - scale independent.
Indeed, the physical value of $ \Lambda_{\overline{MS}} $
cannot be fixed from perturbation theory alone, and needs to be determined
from experiment, which gives $ \Lambda_{\overline{MS}} \simeq 213 MeV$.
In principle, one can solve for $ \Lambda_{\overline{MS}} $ in terms of the 
coupling at any scale, and in particular at the cutoff scale $\Lambda$,
obtaining 
\bea
\Lambda_{\overline{MS}} & = &
\Lambda \, \exp \left ( 
{ - \int^{\alpha_S (\Lambda)} \, {d \alpha_S' \over 2 \, \beta ( \alpha_S') } }
\right )
\nonumber \\
& = & 
\Lambda \, \left ( { \beta_0 \, \alpha_S ( \Lambda ) \over 4 \, \pi } 
\right )^{\beta_1 / \beta_0^2 }
\, e^{ - { 2 \, \pi \over \beta_0 \, \alpha_S (\Lambda ) } }
\, \left [ \, 1 \, + \, O( \alpha_S ( \Lambda) ) \, \right ] \; .
\label{eq:lambda_qcd}
\eea
Not all physical properties can be computed reliably in weak coupling
perturbation theory.
In non-Abelian gauge theories a confining potential 
is found at strong coupling by examining the behavior of the Wilson
loop, 
defined for a large closed loop $C$ as
\beq
\langle \, W( C ) \, \rangle \, = \, 
\langle \, \tr {\cal P} \, \exp \Bigl \{ i g \oint_{C} A_{\mu} (x) dx^{\mu} 
\Bigr \} \, \rangle \;\; ,
\label{eq:wloop_sun}
\eeq
with $A_\mu \equiv t_a A_\mu^a $ and the $t_a$'s the group
generators of $SU(N)$ in the fundamental representation. 
In the pure gauge theory at strong coupling, the leading contribution to
the Wilson loop can be shown to follow an area law for sufficiently
large loops
\beq
\langle \, W( C ) \, \rangle
\; \mathrel{\mathop\sim_{ A \, \rightarrow \, \infty  }} \;
\exp ( - A(C) / \xi^2 ) \; ,
\label{eq:wloop_sun1}
\eeq
where $A(C)$ is the minimal area spanned by the planar loop $C$ \cite{wil74}.
The quantity $\xi $ is the gauge field correlation length, and is
essentially the same [up to a factor $O(1)$] as the inverse of 
$\Lambda_{\overline{MS}}$ in Eq.~(\ref{eq:lambda_qcd}).
The point here is that non-Abelian gauge theories are known to contain a new,
fundamental, dynamically generated length scale, in clear analogy to the
result of Eq.~(\ref{eq:m-cont}) for gravity.
The universal quantity $\xi$ also appears in a number of other physical
observables, including the exponential decay of the Euclidean correlation
function of two infinitesimal loops separated by a distance $|x|$,
\beq
G_{\rm loop-loop} ( x ) \, = \, 
\langle \, \tr {\cal P} \exp \Bigl \{ i g \oint_{C_\epsilon} A_{\mu} (x') dx'^{\mu} 
\Bigr \} (x)
\, 
\tr {\cal P} \exp \Bigl \{ i g \oint_{C_\epsilon} A_{\mu} (x'') dx''^{\mu} 
\Bigr \} (0)
\, \rangle_c \;\; .
\label{eq:box_sun}
\eeq
Here the $C_\epsilon$'s are two infinitesimal loops centered around $x$ ands $0$
respectively, suitably defined on the lattice as elementary square loops, 
and for which one has at sufficiently large separations
\beq
G_{\rm loop-loop} ( x ) 
\; \mathrel{\mathop\sim_{ |x|  \, \rightarrow \, \infty  }} \;
\exp ( - |x| / \xi ) \; .
\label{eq:box_sun1}
\eeq
It is also understood that the inverse of the correlation length $\xi$ 
corresponds to the lowest gauge invariant mass excitation in the gauge theory, 
the scalar glueball with mass $m_0 = 1 / \xi$.
As in the case of gravity [see for comparison Eq.~(\ref{eq:m-invariance})],
the correlation length $\xi$, or equivalently its inverse $m \equiv 1/ \xi$, is
known to be a renormalization group invariant,
\beq
\Lambda \, { d \over d \Lambda } \, m ( \Lambda, g(\Lambda) )
\; = \; 0 \; .
\label{eq:m-invariance-gauge}
\eeq
If one writes $ m \equiv \Lambda \cdot F ( g (\Lambda ) ) $, one obtains immediately
\beq
\beta ( g) \; = \; - { F (g) \over F' (g) } \; ,
\eeq
which then relates, in a direct and explicit way, the dependence of the 
correlation length on the bare
coupling $g$ [through the function $F(g)$] to the Callan-Symanzik beta
function determining the running of the gauge coupling $g$ \cite{wil74}.
The point here is that in gauge theories the emergence
of a nonperturbative length scale can be viewed as a direct consequence
of the RG equations, and of the rather
subtle renormalization group behavior of the gauge coupling 
$\alpha_S $.
It is the combination of these effects that then leads to an entirely
new physical quantum vacuum.

\vskip 30pt

\section{Gravitational Wilson Loop}

\label{sec:loop}

\vskip 10pt

Since the bare cosmological constant can be entirely scaled out of the
theory, the legitimate question arises: how can a non-vanishing
(and indeed small) effective large-scale cosmological constant 
arise out of the field-theoretic treatment of quantum gravity?
The key to this answer lies in the fact that the lattice field
theory itself contains an entirely new dynamically generated scale 
$\xi$, see Eqs.~(\ref{eq:m-cont}) and (\ref{eq:xi_gc}).

To see this, consider elementary parallel transports on the lattice.
Between any two neighboring pair of simplices $s,s+1$ one 
can associate a Lorentz transformation 
${\bf R}^{\mu}_{\;\; \nu} (s,s+1)$, which describes
how a given vector $ V^\mu $ transforms between the local coordinate
systems in these two simplices.
Such a transformation is directly related to the continuum
path-ordered ($P$) exponential of the integral of the local affine connection
$ \Gamma^{\lambda}_{\mu \nu}(x)$ via
\beq
R^\mu_{\;\; \nu} \; = \; \Bigl [ P \; e^{\int_
{{\bf path \atop between \; simplices}}
\Gamma_\lambda d x^\lambda} \, \Bigr ]^\mu_{\;\; \nu}  \;\; ,
\label{eq:rot_cont}
\eeq
with the connection having support on the common interface between the two
simplices.
Next, and in analogy to gauge theories,
one can consider a closed lattice path passing through a large number of
simplices $s$, and spanning a large near-planar closed loop $C$.
Along this closed loop the overall rotation matrix is given by
\beq
R^{\mu}_{\;\; \nu} (C) \; = \;
\Bigl [ \prod_{s \, \subset C} {\bf R}_{s,s+1} \Bigr ]^{\mu}_{\;\; \nu} 
\label{eq:rot_latt}
\eeq
In a semi-classical picture, if the curvature of the manifold is
taken to be small, the expression for the full 
rotation matrix ${\bf R} (C )$ associated with the large near-planar
loop can be re-written in terms of a surface
integral of the large-scale Riemann tensor, projected along the surface
area element bivector $A^{\alpha\beta} (C )$ associated with the loop,
\beq
R^{\mu}_{\;\; \nu} (C) \; \approx \; 
\Bigl [ \, e^{\half \int_S 
R^{\, \cdot}_{\;\; \cdot \, \alpha\beta} \, A^{\alpha\beta} ( C )} \,
\Bigr ]^{\mu}_{\;\; \nu}  \;\; .
\label{eq:rot_riem}
\eeq
Thus, on the one hand, the gravitational Wilson loop provides a way of determining
the effective curvature at large distance scales.

Next consider the case of large metric fluctuations at strong
coupling (large $G$).
The expectation value of the gravitational Wilson loop was defined
in \cite{loop07} as
\beq
\langle \, W(C) \, \rangle  \; = \;
\langle \; \tr [ \, B_C \, 
{\bf R}_1 \; {\bf R}_2 \; ... \; ... \; {\bf R}_n \, ] \; \rangle \;\; ,
\label{eq:wloop}
\eeq
where the ${\bf R}_i$s are the rotation matrices along the path, and
$B_C$ related to a constant bivector characterizing the geometric orientation of
the loop $C$, which again is assumed to be near-planar.
One can then show, by using known properties of the Haar measure
for the rotation group, that, at least for strong coupling and large
area, the Wilson loop follows an area law,
$\langle \, W(C) \, \rangle \,\sim\, \exp \, (- \,{\rm const.}\, A_C ) $.
This last result follows from tiling the interior of the given loop by a 
minimal surface built up of elementary transport loops, in close
analogy to the gauge theory case.
For strong coupling (large $G$) one can write more generally the
result as \cite{loop07}
\beq
\langle \, W(C) \, \rangle \; \mathrel{\mathop\sim_{ A \rightarrow \infty }}
 \, \exp \, ( - \, A_C / {\xi }^2 )
\label{eq:wloop_latt}
\eeq
with $\xi$ determined, by scaling and dimensional arguments,
to be the nonperturbative gravitational correlation length
[see Eq.~(\ref{eq:m-cont})], and again in close analogy to the gauge
theory result of Eq.~(\ref{eq:wloop_sun1}).
In the following we shall assume, in close analogy to what is known
to happen in non-Abelian gauge theories due to scaling, that even though the above form
for the gravitational Wilson loop was derived in the extreme strong coupling
limit, it will remain valid throughout the whole strong coupling
phase and all the way up to the nontrivial ultraviolet fixed point,
with the correlation length $\xi \rightarrow \infty $ the only 
relevant, and universal, length scale in the vicinity of such
a fixed point.
The evidence for the existence of such an UV fixed point comes
from three different sources, 
which have recently been reviewed, for example, in \cite{book}
and references therein.
In the gravitational Wilson loop result of Eq.~(\ref{eq:wloop_latt}),
$\xi$ is therefore identified with the renormalization group invariant 
quantity obtained by integrating the $\beta$-function for the 
Newtonian coupling $G$, see Eqs.~(\ref{eq:m-cont}) and (\ref{eq:xi_gc}).

One can now compare the quantum result at strong
coupling, Eq.~(\ref{eq:wloop_latt}), with the semiclassical result
that follows from Eq.~(\ref{eq:rot_riem}).
The latter gives
\beq
W(C) \, \sim \, \Tr \left ( B_C \, \exp \, 
\left \{ \, \half \,
\int_{S(C)}\, R^{\, \cdot}_{\;\; \cdot \, \mu\nu} \, A^{\mu\nu}_{C} \; 
\right \} \right ) \; .
\label{eq:wloop_curv}
\eeq
where again $B_C$ is constant bivector characterizing the orientation 
of the near-planar loop $C$.
Then for a smooth background classical manifold with
constant or near-constant large-scale curvature,
\beq
R_{\mu\nu\lambda\sigma} = \third \, \lambda \, ( 
g_{\mu\nu} \, g_{\lambda\sigma} \, - \, 
g_{\mu\lambda} \, g_{\nu\sigma} )
\eeq
one immediately obtains from the identification of the area terms
in the two Wilson loop expressions
the following result for the average semi-classical 
curvature at large scales
\beq
{\bar R} \; \sim \; + \, 1 / \xi^2  \; .
\label{eq:xi_r}
\eeq 
Note that a key ingredient in the derivation is the fact that
both in the quantum result of Eq.~(\ref{eq:wloop_latt}) and 
in the semi-classical result of Eq.~(\ref{eq:wloop_curv}) 
the exponent contains the {\it area} of the loop.
An equivalent way of phrasing the statement of Eq.~(\ref{eq:xi_r})
uses the classical field equations in the absence of matter, 
$R = 4 \lambda$.
The latter suggests one should view $1 / \xi^2$,
up to a constant of proportionality of order one, 
as the observed scaled cosmological constant,
\beq
\third \, \lambda_{obs} \; \simeq \;  + \, { 1 \over \xi^2 }  \; .
\label{eq:xi_lambda}
\eeq 
This last quantity can then be considered as a measure of 
the gravitational vacuum energy, in analogy to the (by now well established) 
non-Abelian gauge theory vacuum condensate result,
$ \langle F_{\mu\nu}^2 \rangle \, \simeq \, 1 / \xi^4 $,
whose gravity analog can be written, equivalently, as
\beq
\langle \, R \, \rangle \, \propto \, { 1 \over \xi^2  } \; .
\eeq
The nonperturbative treatment
of lattice quantum gravity has thus added one more ingredient
to the puzzle: while the bare cosmological constant $\lambda_0$ can be
completely scaled out of the problem,
a new RG invariant scale $\xi$ of Eq.~(\ref{eq:m-cont}) appears, and
is here identified with the {\it effective} cosmological constant.
The fact that the RG invariant quantity $\xi$ presents a natural
candidate for the observed cosmological constant was proposed
independently in \cite{per99}, one of the first papers
to bring up the cosmological consequences of such an
identification.

\vskip 30pt

\section{Effective Field Equations}

\label{sec:eff}

\vskip 10pt

An important physical consequence, implied by the identification of 
the RG invariant $\xi$ in Eq.~(\ref{eq:m-cont}) with the effective,
long distance $1/ \sqrt{ \lambda }$, is that one expects (as in gauge
theories) $\xi$ to determine the scale dependence of the effective 
Newton's constant $G$ appearing in the field equations.
The latter is a solution of the renormalization group equations for $G$, given in 
Eqs.~(\ref{eq:grun_k}), (\ref{eq:m-cont}) and (\ref{eq:xi_gc}).
Specifically, if one follows Eq.~(\ref{eq:grun_k}), one obtains a
momentum-dependent $G(k)$.
This needs to be reexpressed in a covariant way, so that effects 
from it can be computed consistently for general problems,
involving arbitrary background geometries.
The first step in analyzing the consequences of a running of $G$
is therefore to rewrite the expression for $G(k)$ in a 
manifestly coordinate-independent way.
This can be done either by the use of a nonlocal Vilkovisky-type 
effective gravitational action \cite{vil84,bar85,ven90,bar90,bar03},
or by the use of a set of consistent effective field equations \cite{hw05}.
In either case one goes from momentum to position space by applying
the prescription $k^2 \rightarrow - \Box$.
This then gives for the quantum-mechanical running of the gravitational
coupling the replacement
\beq
G  \;\; \rightarrow \;\; G( k ) \;\; \rightarrow \;\; G( \Box ) \; .
\eeq
As a consequence, the running of $G$ in the vicinity of the UV fixed
point is of the form
\beq
G ( \Box ) \, = \, G_0 \left [ \; 1 \, 
+ \, c_0 \left ( { 1 \over - \xi^2 \, \Box  } \right )^{1 / 2 \nu} \, 
+ \, \dots \, \right ] \; ,
\label{eq:grun_box}
\eeq
where $\Box \equiv g^{\mu\nu} \nabla_\mu \nabla_\nu$ is the covariant
d'Alembertian, and the dots represent higher
order terms in an expansion in $1 / ( \xi^2 \, \Box ) $.
Note that $G_0 \equiv G_c $ in the above expression should be identified to a first
approximation with the laboratory scale value of Newton's constant, 
$ \sqrt{G_c} \sim 1.6 \times 10^{-33} cm$, whereas
$\xi \sim 1 / \sqrt{\lambda/3} \sim 1.51 \times 10^{28} {\rm cm} $.
Current numerical evidence from Euclidean lattice gravity gives
$c_0 \simeq 8.0 > 0$ (implying infrared growth) and $\nu = \third $ \cite{ham00}.

It is worth mentioning here that one could consider an infrared
regulated version of $G(\Box)$, where the infrared cutoff 
$\mu \sim \xi^{-1}$ is introduced, so that in Fourier space 
$k > \xi^{-1}$ and thus spurious infrared divergences at small $k$ are removed.
This can be achieved by the (QCD renormalon-inspired) replacement
$k^2 \rightarrow k^2 + m^2 $ in Eq.~(\ref{eq:grun_k})
with $m= 1/ \xi $ as the infrared cutoff.
In position space this then leads to the IR regulated form
of Eq.~(\ref{eq:grun_box})
\beq
G ( \Box ) \, = \, G_0 \left [ \, 1 \, 
+ \, c_0 \left ( { 1 \over - \, \xi^2 \, \Box  + 1 } \right )^{1 / 2 \nu} 
\, \right ] \; .
\label{eq:grun_box_reg}
\eeq
Nevertheless, in the following it will be adequate to just consider the expression
in Eq.~(\ref{eq:grun_box}), although most of the discussion
given below is quite general, and does not hinge on this specific choice.

One possible approach to develop an effective theory
is then to write down a set of classical effective, but
nonlocal, field equations of the form
\beq
R_{\mu\nu} \, - \, \half \, g_{\mu\nu} \, R \, + \, \lambda \, g_{\mu\nu}
\; = \; 8 \pi \, G( \Box  )  \; T_{\mu\nu}
\label{eq:field_box}
\eeq
with $\lambda \simeq 3 / \xi^2  $ and $ G( \Box )$ given above,
and a strong nonlocality from the $G(\Box)$ term.
From this the running of $G$ can then be worked out in detail for specific coordinate choices.
For example, in the static isotropic case one finds 
a gradual slow rise in $G$ with distance
\beq
G \; \rightarrow \; G(r) \; = \; 
G \, \left ( 1 \, + \, 
{ c_0 \over 3 \, \pi } \, { r^3 \over \xi^3 } 
\, \ln \, { \xi^2 \over  r^2 }  
\, + \, \dots
\right )
\label{eq:g_small_r3}
\eeq
in the regime $r \gg 2 \, M \, G$ with $2MG$ is the horizon radius \cite{hw06}.

To aid in the interpretation of the physical content
of the theory, one notes that the nonlocal effective field equation of 
Eq.~(\ref{eq:field_box}) can be recast in a form very similar to the classical
field equations, but with an additional source term coming
from the vacuum polarization contribution \cite{ht10}.
For this purpose it is useful to decompose the full source term in the effective
field equations by first writing
\beq
G(\Box) = G_0 \, \left ( 1 \, +  {\delta G(\Box) \over G_0} \right ) 
\;\;\;\;\;\;  {\rm with} \;\;\;\;\;
{\delta G(\Box) \over G_0} \equiv 
c_0 \left ( { 1 \over - \xi^2 \, \Box } \right )^{1 / 2 \nu}  \; .
\label{eq:grun_box_1}
\eeq
Then the full source term can be written as a sum of two parts,
\beq
\left ( 1 + {\delta G(\Box) \over G_0}  \right ) \, T_{\mu\nu}  \, = \,
T_{\mu\nu}   +  T_{\mu\nu}^{vac} \; .
\label{eq:tmunu_tot}
\eeq
The second, vacuum part involves the nonlocal term
\beq
T_{\mu\nu}^{vac} \, \equiv \,  {\delta G(\Box) \over G_0} \; T_{\mu\nu} \; .
\label{eq:tmunu_vac}
\eeq
with the covariant d'Alembertian operator 
$ \Box \; = \; g^{\mu\nu} \, \nabla_\mu \nabla_\nu  $
acting here on the second rank tensor $T_{\mu\nu}$,
\bea
\nabla_{\nu} T_{\alpha\beta} \, = \, \partial_\nu T_{\alpha\beta} 
- \Gamma_{\alpha\nu}^{\lambda} T_{\lambda\beta} 
- \Gamma_{\beta\nu}^{\lambda} T_{\alpha\lambda} \, \equiv \, I_{\nu\alpha\beta}
\nonumber
\eea
\beq 
\nabla_{\mu} \left ( \nabla_{\nu} T_{\alpha\beta} \right )
= \, \partial_\mu I_{\nu\alpha\beta} 
- \Gamma_{\nu\mu}^{\lambda} I_{\lambda\alpha\beta} 
- \Gamma_{\alpha\mu}^{\lambda} I_{\nu\lambda\beta} 
- \Gamma_{\beta\mu}^{\lambda} I_{\nu\alpha\lambda}  \; ,
\label{eq:box_on_tensors}
\eeq
In this picture, therefore, the running of $G$ can be viewed 
as contributing to a sort of vacuum fluid, introduced in order to account for 
the gravitational quantum vacuum-polarization contribution.
Consistency of the full covariant, nonlocal field equations then
requires that the sum of the two $T_{\mu\nu}$ contributions
be conserved,
\beq
\nabla^\mu \left ( T_{\mu\nu}   +  T_{\mu\nu}^{vac}  \right ) = 0 \; ,
\label{eq:bianchi}
\eeq
in consideration of the contracted Bianchi identity 
satisfied by the Ricci tensor.
Due to the appearance, in $G(\Box)$ of Eq.~(\ref{eq:grun_box}),
of the inverse of the covariant Laplacian raised to a 
fractional power, it seems wise to consider a regulated version
that can be used reliably for practical calculations.
One possibility is to compute the effect of $\Box^n$ for positive
integer $n$, and then analytically continue the results to 
$n \rightarrow -1/2\nu$, as was done in \cite{hw05}.
Alternatively, $G(\Box)$ can be defined via a regulated parametric integral 
representation \cite{lop07}.
In view of the discussion to follow, it will be advantageous to
write the relevant nonlocal part of $G(\Box)$ as
\beq
\left ( { 1 \over - \, \Box \, (g) + \mu^2 } \right )^{1/ 2 \nu } 
\, = \, 
{ 1 \over \Gamma ( { 1 \over 2 \nu } ) } \,
\int_0^\infty d \alpha \; \alpha^{  1 / 2 \nu - 1 } \;
e^{  - \alpha \, [ - \Box (g) + \mu^2 ] } \; ,
\label{eq:gbox_exp}
\eeq
where $\mu \rightarrow 0 $ is a suitable infrared regulator.

Next consider what happens in the case of a running cosmological constant
entering the effective field equation of Eq.~(\ref{eq:field_box}).
Earlier in this work we discussed the fact that a running
cosmological constant $\lambda (k) $ is both inconsistent with the overall scaling
properties of the gravitational functional integral
in the continuum and on the lattice [see Secs.~(\ref{sec:cont})
and (\ref{sec:latt})], and with gauge invariance in the 
perturbative treatment about two dimensions [Sec.~(\ref{sec:pert})].
The expectation is therefore that serious inconsistencies will arise
when a running cosmological constant is formulated
within a fully covariant effective theory approach.
The first step is therefore to promote again an RG running
in momentum space to a manifestly covariant form,
$\lambda (k) \rightarrow \lambda ( \Box ) $ in the effective
field equation of Eq.~(\ref{eq:field_box}).
To be more specific, consider the case of a scale dependent
$\lambda (k)$, which we will write here as 
$\lambda = \lambda_0 + \delta \lambda (k) $.
We will also assume, again for concreteness, that 
$\delta \lambda (k) \sim c_1 (k^2)^{-\sigma}$,
where $c_1$ and $\sigma$ are some constants.
Then make again the transition to coordinate space 
by replacing $ k^2 \rightarrow - \Box $.
This leads to
\beq
\delta \lambda ( \Box ) \; \sim \; ( - \Box (g) + \mu^2 )^{- \sigma}
\; ,
\eeq
where we have been careful and used again the infrared 
regulated expression given in Eq.~(\ref{eq:gbox_exp}).
The effective field equations in Eq.~(\ref{eq:field_box})
then contain the following additional 
(running cosmological) term
\beq
\delta \lambda (\Box) \cdot g_{\mu\nu}  \; = \;
\, c_1 \; { 1 \over \Gamma ( \sigma ) } \,
\int_0^\infty d \alpha \; \alpha^{  \sigma - 1 } \;
e^{  - \alpha \, ( - \Box (g) + \mu^2 ) } \cdot g_{\mu\nu}  
\; = \;  \, c_1 \, ( \mu^2 )^{- \sigma} \cdot g_{\mu\nu}  \; .
\eeq
The result therefore is still a numerical constant multiplying the metric
$g_{\mu\nu}$.
Use has been made here of the key result that covariant derivatives
of the metric tensor vanish identically,
\beq
\nabla_\lambda \, g_{\mu\nu} \; = \; 0 \; .
\label{eq:der_met}
\eeq
The conclusion of this exercise is therefore that $\lambda$ cannot run.
Note also another key aspect of the derivation: what matters is
not just the form of $\lambda (\Box)$, but also the object it
acts on.
This last aspect is missed completely if one just focuses on $\lambda (k)$.
Moreover, the above rather general argument applies also 
to possible additional contributions to the vacuum energy
from various condensates
and nonzero vacuum expectation values of matter fields, such as the 
QCD color field condensate, the quark condensate and the Higgs field.
One is lead therefore to the conclusion that, quite generally, a
running of $\lambda$ in the
effective field equations inevitably ends up in conflict 
with general covariance, in essence by virtue of Eq.~(\ref{eq:der_met}).

\vskip 30pt

\section{Effective Action}

\label{sec:action}

\vskip 10pt

The previous section discussed how the RG running of $G (\Box) $ can be
incorporated in a set of manifestly covariant effective field
equations.
It was also shown that a running of the cosmological constant
in the same equations is essentially ruled out by the requirement 
of general covariance.
One main advantage of Eq.~(\ref{eq:field_box}) is that it is actually
tractable, and leads to a number of reasonably unambiguous predictions
for homogeneous isotropic and static isotropic background metrics \cite{hw06}.

In this section we will approach the same problem from a slightly
different perspective, namely from the point of view
of an effective gravitational action.
In view of the discussion presented earlier in this work, it should
be clear that such an effective action will depend on the
two renormalized, dimensionful parameters $G$ and $\xi$.
Note that we will focus here mostly on the case of pure gravity, as the
addition of matter will leave most of the main conclusions unchanged
(as was the case in the previous section).
Within the framework of an effective action approach,
the running of the coupling constants can be implemented
by the use of a manifestly covariant effective gravitational action 
\cite{vil84,bar85,ven90,bar90,bar03}.
First consider the cosmological term, for
which we write again
\beq
\lambda_0  \;\; \rightarrow \;\;  \lambda_0 ( k ) \;\; \rightarrow
\;\; \lambda_0 ( \Box ) \; .
\eeq
It is then easy to see that
\beq
\lambda_0 \int d^4 x \, \sqrt{g} \;\; \rightarrow \;\;
\int d^4 x \, \sqrt{g} \;\, \lambda_0 ( \Box ) \cdot 1 
\eeq
is meaningless, as $ \lambda_0 ( \Box ) $ has nothing to act on.
Therefore the $\lambda_0$ term in the gravitational action cannot 
be made to run, no matter how hard one tries.
The implication again here is that if $\lambda_0$ is somehow made to 
run, this can only be achieved by an explicit breaking of 
general covariance.
\footnote{
Of course one could {\it force} $\lambda_0$ to run by writing for the
integrand $f(R)^{-1} A(\Box)R f(R)$ where $f$ is some arbitrary
function of the scalar curvature, or of any other quantum field for that matter. 
But to us this procedure seems entirely artificial.}

One further notices that this is clearly {\it not} the case for the
rest of the gravitational action, and in particular
for the running of $G$, as given in Eq.~(\ref{eq:grun_box}).
Indeed, consider here for concreteness the following nonlocal 
effective gravitational action
\beq
I \; = \; - \, { 1 \over 16 \pi \, G } \int d^4 x \sqrt{g} \; \sqrt{R} \,
\left( 1 \, - \, A (\Box) \right ) \sqrt{R}
\label{eq:eff_action}
\eeq
with [see Eq.~(\ref{eq:grun_box})]
\beq
A (\Box) \; \equiv \;  
c_0 \, \left ( - \xi^2 \, \Box \right )^n
\label{eq:a_box}
\eeq
and $G$ a true constant.
In last expression $n$ is taken to be an integer, with $n \rightarrow - 1 / 2 \nu $
at the end of the calculation.
The next step is to compute the variation of the above effective action.
Note that another possibility would have been to have the $G(\Box)$ act on the
matter term, $\half \int \sqrt{g} \; G(\Box) \, g^{\mu\nu} \, T_{\mu\nu}$,
but we will not pursue this possibility here.
The expression inside the integral requires the evaluation
of four separate variation terms,
\bea
& - & \half \; \sqrt{g} \; \delta g^{\mu\nu} \, g_{\mu\nu} \, \sqrt{R} \, 
\left( 1 \, - \, A (\Box) \right ) \sqrt{R}
\, + \, \sqrt{g} \, \delta \sqrt{R} \, 
\left( 1 \, - \, A (\Box) \right ) \sqrt{R}
\nonumber \\
& - & n \, \sqrt{g} \; \sqrt{R} \; { A (\Box) \over \Box } \, 
(\delta \, \Box) \, \sqrt{R}
\, + \, \sqrt{g} \; \sqrt{R} \; 
\left( 1 \, - \, A (\Box) \right ) \, \delta \sqrt{R}  \; .
\label{eq:variation}
\eea
These in turn require the following elementary variations,
\beq
\delta \sqrt{g}  \; = \; - \, \half \, \sqrt{g} \, g_{\mu\nu} \, \delta g^{\mu\nu}
\eeq
and 
\beq
\delta R \; = \; g^{\mu\nu} \, \delta R_{\mu\nu} \, + \, 
R_{\mu\nu} \, \delta g^{\mu\nu}
\label{eq:delta_r}
\eeq
with 
\beq
\delta R_{\mu\nu} \; = \; 
\nabla_{\alpha} \left ( \delta \Gamma^{\alpha}_{\;\; \mu\nu} \right )
\, - \, 
\nabla_{\mu} \left ( \delta \Gamma^{\alpha}_{\;\; \alpha\nu} \right )
\; ,
\eeq
for which one needs
\beq
\delta \Gamma^{\alpha}_{\;\; \mu\nu} \; = \;  \half \, g^{\alpha\beta}
\, \left [ 
\nabla_{\mu} \, \delta g_{\beta\nu} \, + \,
\nabla_{\nu} \, \delta g_{\beta\mu} \, - \,
\nabla_{\beta} \, \delta g_{\mu\nu} \right ] \; .
\label{eq:gammavar}
\eeq
It then follows that in Eq.~(\ref{eq:delta_r}) 
\beq
g^{\mu\nu} \, \delta R_{\mu\nu} \; = \; 
\nabla_{\mu} \nabla_{\nu} \left (
\, - \, \delta g^{\mu\nu} \, + \, 
g^{\mu\nu} \, g_{\alpha\beta} \, \delta g^{\alpha\beta} \right )
\; = \; 
g_{\alpha\beta} \, \Box \, \delta g^{\alpha\beta}
\, - \, 
\nabla_{( \mu} \nabla_{\nu )} \, \delta g^{\mu\nu} \; ,
\eeq
which gives the second and last terms in Eq.~(\ref{eq:variation}).
Use has been made here of $\delta g_{\mu\nu} = - g_{\mu\alpha} \,
g_{\nu\beta} \, \delta g^{\alpha\beta} $.
Note that in general $\Box \, \nabla_\mu \neq \nabla_\mu \Box$, and that
$\Box \, g_{\mu\nu} =0 $ but $\Box \, \delta g_{\mu\nu} \neq 0 $.
For the variation of the covariant d'Alembertian 
\beq
\delta ( \Box ) \; = \; \delta g^{\mu\nu} \, \nabla_\mu \nabla_\nu 
\, - \, g^{\mu\nu} \, \delta \Gamma_{\mu\nu}^{\sigma} \, \nabla_{\sigma}
\eeq
one needs the variation of $\Gamma_{\mu\nu}^{\sigma}$ given in 
Eq.~(\ref{eq:gammavar}), which leads to
\beq
\delta ( \Box ) \; = \; 
\delta g^{\mu\nu} \, \nabla_\mu \, \nabla_\nu 
\, + \, ( \nabla_\mu \, \delta g^{\mu\nu} ) \, \nabla_\nu 
\, - \, \half \, g_{\mu\nu} \, g^{\alpha\beta} \, 
( \nabla_\beta \, \delta g^{\mu\nu} ) \, \nabla_\alpha \;\; .
\eeq
Generally one encounters expression that need to be 
properly symmetrized, as in the case of 
\beq
\delta ( \Box^n ) \, \rightarrow \, \sum_{k=1}^n \, \Box^{k-1}
\, (\delta \, \Box) \, \Box^{n-k} \; .
\label{eq:delta_a}
\eeq
Eventually this leads to rather lengthy and complicated expressions;
although these can be worked out in detail,
in the following we will first consider, by choice, only one 
such ordering, namely 
$ \delta ( \Box^n ) \, \rightarrow \, n \, \Box^{n-1} \, \delta (\Box ) $,
which then gives simply $\delta (A) = n [ A(\Box) / \Box ] \,  \delta (\Box) $.
Several integrations by parts need to be performed next, involving both 
$\Box^n$ (with integer $n$) and $ g_{\mu\nu} \, \Box - \nabla_{( \mu} \nabla_{\nu )} $,
required in order to isolate the $\delta g^{\mu\nu}$ term.
In general one has to be careful about the ordering of covariant
derivatives, whose commutator is non-vanishing
\beq
[ \nabla_{\mu} , \nabla_{\nu} ] \, 
T^{\alpha_1 \, \alpha_2 \dots}_{\;\;\;\;\;\;\;\;\;\;\;\; \beta_1 \, \beta_2 \dots}
\; = \; - \sum_i \, R_{\mu\nu\sigma}^{\;\;\;\;\;\;\; \alpha_i} \,
T^{\alpha_1 \dots \sigma \dots}_{\;\;\;\;\;\;\;\;\;\;\;\;\; \beta_1 \dots}
- \sum_j \, R_{\mu\nu\beta_j}^{\;\;\;\;\;\;\;\;\; \sigma} \,
T^{\alpha_1 \dots}_{\;\;\;\;\;\;\; \beta_1 \dots \sigma \dots}
\eeq
with the $\sigma$ index in $T$ in the $i$-th position in the first
term, and in the $j$-th position in the second term.
As a consequence, the $O(R)$ commutator terms generally give
rise to higher derivative terms in the effective field equations, due
to the fact that the zeroth order terms in the action are already $O(R)$.
After all these manipulations, the effective field equations for zero
cosmological constant take on the form
\bea
\left ( R_{\mu\nu} \, - \, \half \, g_{\mu\nu} \, R \right ) \,
\left ( 1 \, - \, { 1 \over \sqrt{R} } \, A(\Box) \, \sqrt{R} \right )
\nonumber
\eea
\bea
\, - \, \left ( g_{\mu\nu} \, \Box \, - \, \nabla_{( \mu} \nabla_{\nu )}
\right ) 
\, \left ( { 1 \over \sqrt{R} } \, A(\Box) \, \sqrt{R} \right )
\, + \, n \, 
\left ( \nabla_\mu \, { A(\Box) \over \Box } \, \sqrt{R} \right ) \,
\left ( \nabla_\nu \, \sqrt{R} \right ) \,
\nonumber
\eea
\bea
- \, \half \, n \, g_{\mu\nu} \left \{
\left ( \nabla^\sigma \, { A(\Box) \over \Box } \, \sqrt{R} \right ) \,
\left ( \nabla_\sigma \, \sqrt{R} \right ) 
\, + \, 
\left ( { A(\Box) \over \Box } \, \sqrt{R} \right ) \,
\left ( \Box \, \sqrt{R} \right )  \right \}
= 8 \pi G \, T_{\mu\nu} \; .
\label{eq:field_var}
\eea
Note that the above effective field equations are not symmetric in 
$\mu \leftrightarrow \nu$ due to our specific choice of operator ordering.
Note also that taking the covariant divergence of the l.h.s is expected to give zero,
which is required for consistency of the field equations
(for some terms it is clear that they give zero by inspection).
Unfortunately the above effective field equations are still rather
complicated.
Note though that generally any terms of $O(R^2)$ can
safely be dropped, if one is interested in the long distance,
small curvature limit.
For completeness, we quote here the result for arbitrary operator ordering, as
in Eq.~(\ref{eq:delta_a}), where
the generic term has the form $\Box^{k-1} \, \delta ( \Box ) \, A ( \Box )
\, \Box^{-k} $ with $k = 1 \dots n $.
In this case the last two terms on the l.h.s. of
Eq.~(\ref{eq:field_var}) become
\bea
\, + \, 
\left ( \nabla_\mu \, \Box^{k-1} \, \sqrt{R} \right ) \,
\left ( \nabla_\nu \, { A(\Box) \over \Box^k } \, \sqrt{R} \right ) \,
\nonumber
\eea
\bea
- \, \half \, g_{\mu\nu} \left \{
\left ( \nabla^\sigma \, \Box^{k-1} \, \sqrt{R} \right ) \,
\left ( \nabla_\sigma \, { A(\Box) \over \Box^k } \, \sqrt{R} \right ) \,
\, + \, 
\left ( \Box^{k-1} \, \sqrt{R} \right ) \,
\left ( \Box \, { A(\Box) \over \Box^k } \, \sqrt{R} \right ) \,
 \right \} \; .
\label{eq:field_var_k}
\eea
The full effective field equations are then obtained by summing over
$k$, as in Eq.~(\ref{eq:delta_a}).

One more possibility is to generalize the effective action in 
Eq.~(\ref{eq:eff_action}) to the form 
\beq
I \; = \; - { 1 \over 16 \pi \, G } \int d^4 x \sqrt{g} \, R^{1-\alpha}
\left( 1 \, - \, A (\Box) \right ) R^{\alpha} \; ,
\eeq
which now depends on a parameter $\alpha$ taking values
between zero and one; the previous case
then corresponds to the symmetric choice $\alpha=1/2$.
This last action can also be symmetrized between the first and
second curvature terms, but we shall not pursue that here in order
to keep things simple.
Following the same procedure that lead to Eq.~(\ref{eq:field_var}),
one obtains for the field equations with zero cosmological constant
the following expression
\bea
R_{\mu\nu} \, - \, \half \, g_{\mu\nu} \, R \,
\, + \, \half \, g_{\mu\nu} \, R \, \cdot \, { 1 \over R^\alpha } \, A(\Box) \, R^{\alpha}
\nonumber
\eea
\bea
\, - \, R_{\mu\nu} \, 
\left \{ (1-\alpha) \, { 1 \over R^{\alpha} } \, A(\Box) \, R^{\alpha} 
\, + \, \alpha \, { 1 \over R^{1-\alpha} } \, A(\Box) \, R^{1-\alpha} \right \}
\nonumber
\eea
\bea
\, - \, 
\left ( g_{\mu\nu} \, \Box \, - \, \nabla_{( \mu} \nabla_{\nu )}
\right ) 
\left \{ (1-\alpha) \, { 1 \over R^{\alpha} } \, A(\Box) \, R^{\alpha} 
\, + \, \alpha \, { 1 \over R^{1-\alpha} } \, A(\Box) \, R^{1-\alpha}
\right \}
\nonumber
\eea
\bea
\, + \, n \, \left ( \nabla_\mu  \, { A(\Box) \over \Box } \, R^{1-\alpha} \right ) \,
\left (  \nabla_\nu \, R^{\alpha} \right )
\nonumber
\eea
\bea
- \, \half \, n \, g_{\mu\nu} \left \{
\left ( \nabla^\sigma \, { A(\Box) \over \Box } \, R^{1-\alpha} \right ) \,
\left ( \nabla_\sigma \, R^\alpha \right ) 
\, + \, 
\left ( { A(\Box) \over \Box } \, R^{1-\alpha} \right ) \,
\left ( \Box \, R^\alpha \right )  \right \}
= 8 \pi G \, T_{\mu\nu}
\label{eq:field_var1}
\eea
which incidentally shows that the choice of either $\alpha=1$ or $\alpha=0$ 
is problematic.

Then one final technical question remains, namely what is the relationship between
the above effective field equations 
[Eq.~(\ref{eq:field_var}) or Eq.~(\ref{eq:field_var1}] and the
clearly more economical field equations given earlier in Eq.~(\ref{eq:field_box}).
Obviously, the equations obtained here from a variational principle
are significantly more complicated.
They contain a number of non-trivial terms, some of which are
reminiscent of the $1 \, + \, A (\Box)$ term in
Eq.~(\ref{eq:field_box}),
and others with a rather different structure (such as the 
$ g_{\mu\nu} \, \Box \, - \, \nabla_{( \mu} \nabla_{\nu )} $ term).
Note that many of the new additional curvature terms that appear on the l.h.s. 
of the effective field equations in Eqs.~(\ref{eq:field_var}) 
and (\ref{eq:field_var1}) can be moved, equivalently, to the r.h.s..
To do so, one makes use of the fact that to zeroth order in the quantum 
correction proportional to $A(\Box)$ 
\beq
R \; = \; 
4 \, \lambda \, - \, 8 \pi \, G \, T^{\lambda}_{\;\;\lambda} \; ,
\eeq
which then allows, again, a separation of the source term
in pure matter ($T_{\mu\nu}$) and vacuum
polarization ($T^{vac}_{\mu\nu}$) contributions, as was done earlier in
Eq.~(\ref{eq:tmunu_tot}).
Generally, some of the issues that come up in comparing
effective field equations reflect an ambiguity
of where, and at what stage, the replacement $G \rightarrow G(\Box)$
is performed.
Nevertheless one would hope that when asking the right
physical question the answer would largely be unambiguous.
It is of course possible that, when restricted to specific metrics such
as the Robertson-Walker one and its perturbations, the two sets of 
effective field equations will ultimately give similar results,
but in general this remains still a largely open question.
One possibility is that both sets of field equations
describe the same running of the gravitational coupling, 
up to curvature squared (higher derivative) terms, which then
become irrelevant at very large distances.
In any case, the main purpose of our exercise here
was to show that in either case [via Eq.~(\ref{eq:field_box}) 
or Eq.~(\ref{eq:field_var})] the running
of $G (\Box)$ clearly leads to non-vanishing effects which 
are non-trivial.

\vskip 30pt

\section{Renormalization via Continuum Truncation Methods}

\label{sec:trunc}

\vskip 10pt

A number of approximate continuum renormalization group methods have 
been developed, which can be used to construct RG flows and thus
estimate the scaling exponents.
Let us mention here one example, as an illustration
for the kind of rather delicate issues that might arise.
An approach closely related to the $2+\epsilon$ 
perturbative expansion
for gravity discussed earlier is the derivation of approximate RG flow equations
from the changes of the effective action with respect
to an infrared cutoff $\mu$.
In some ways the method is a variation of Wilson's original
momentum slicing technique, originally developed to obtain approximate
renormalization group recursions for the couplings.
As an example, in the case of a scalar field
one starts from the partition function
\beq
\exp ( W[J] ) \; = \; \int [ d \phi ] \, \exp \left \{
\, - \, \half \phi \cdot C^{-1} \cdot \phi 
\, - \, I_{\Lambda} [ \phi ] + J \cdot \phi \right \} \; ,
\eeq
where the $ C ( k, \mu )$ term introduces an 'infrared cutoff term'. 
In order for it to act as an infrared cutoff, it needs to be small
for $k<\mu$, tending to zero as $k \rightarrow 0$,
with $k^2 C(k,\mu)$ large when $k > \mu$.
Since the method is useful in the vicinity of
the fixed point, where physical relevant scales are
much smaller than the ultraviolet cutoff $\Lambda$, it is argued that
the detailed nature of this cutoff is not too relevant.
Taking a derivative of the function $W[J]$ with respect to $\mu$ gives
\beq
{ \partial W[J] \over \partial \mu } \; = \;
\, - \, \half \left [ \,
{ \delta W \over \delta J } \cdot { \partial C^{-1} \over \partial \mu }
\cdot  { \delta W \over \delta J }
\, + \, \tr \left ( { \partial C^{-1} \over \partial \mu }
{ \delta^2 W \over \delta J \, \delta J } \right ) \right ] \; .
\eeq
The latter can then be expressed in terms of the Legendre transform
$\Gamma [ \phi ] = - W [J] - \half \phi \cdot C^{-1} \cdot \phi + J \cdot \phi $
as
\beq
{ \partial \, \Gamma [\phi] \over \partial \mu } \; = \; 
- \, \half \tr \left [ \, { 1 \over C } \,  
{ \partial C \over \partial \mu }
\cdot  \left ( 1 \, + \, C \cdot { \delta^2 \Gamma \over \delta \phi \, \delta \phi }
\right )^{-1} \right ] \; .
\label{eq:gamma-rg}
\eeq
Here $\phi \equiv \delta W / \delta J$ is regarded as the classical field,
and the traces are later simplified by going to momentum space.
One issue that needs to be settled in advance is the choice for the
cutoff function $C(k, \mu)$.
Then given this choice one computes the effective action 
$\Gamma [\phi]$ in a derivative expansion containing terms 
$\partial^n \phi^m$ and calculable $\mu$-dependent coefficients.
It is also customary to write the cutoff function as
$C(k,\mu)= \mu^{\eta-2} C(k^2 / \mu^2 )$, so as to anticipate
an anomalous dimensions ($\eta$) for the $\phi$ field,
and assume for the remaining function (now of a single variable) that
$C(q^2)=q^{2p}$ with $p$ a non-negative integer \cite{mor94}.

In the gravitational case one proceeds more or less in a similar way.
First note that the gravity analog of Eq.~(\ref{eq:gamma-rg}) is
\beq
{ \partial \, \Gamma [g] \over \partial \mu } \; = \;
- \half \tr \left [ \,
{ 1 \over C }  
{ \partial C \over \partial \mu }
\cdot  \left ( 1 \, + \, C \cdot 
{ \delta^2 \Gamma \over \delta g \, \delta g }
\right )^{-1} \right ] 
\label{eq:gamma-rg-grav}
\eeq
where $g_{\mu\nu} \equiv \delta W / \delta J_{\mu\nu}$ is 
the classical metric.
The effective action then contains an Einstein and a cosmological term
\beq
\Gamma_\mu [g] \; = \; - { 1 \over 16 \pi \, G(\mu) } \int d^d x \sqrt{g} 
\, [ \, R(g) \, - \, 2 \lambda ( \mu ) \, ] + \dots \; ,
\eeq
as well as gauge fixing and possibly higher derivative terms 
\cite{reu98,lit04}.
After the addition of a suitable background harmonic gauge fixing term 
with gauge parameter $\alpha$, the choice of a (scalar) cutoff function
is required, $C^{-1} (k,\mu) = (\mu^2 - k^2) \theta ( \mu^2 - k^2 )$
[for more details see for example \cite{lit04}].
The latter is then inserted into the path integral
\beq
\int [ d h ] \, \exp \left \{
- \half \, h \cdot C^{-1} \cdot h
\, - \, I_{\Lambda} \, [ g ] \, + \, J \cdot h \right \} \; .
\eeq
Note that the additional momentum-dependent cutoff term violates both the
weak field general coordinate invariance, as well as the general
rescaling invariance of Eq.~(\ref{eq:rescale}).
Subsequently the solution of the resulting renormalization group
equation for the two couplings $G(k)$ and $\lambda (k)$ 
is truncated to the Einstein
and cosmological terms, a procedure which is more or less equivalent to the
derivative expansion discussed previously for the scalar case.
A nontrivial fixed point in the couplings
($G^{*}, \lambda^{*}$) is then found, 
generally with complex eigenvalues $\nu^{-1}$,
and some dependence on the gauge parameters \cite{lau02}.

There seem to be two problems with the above approach (apart from
the reliability and convergence of the truncation procedure, 
which is an entirely separate issue).
The first problem is an explicit violation of the scaling properties of
the gravitational functional integral, see
Eqs.~(\ref{eq:rescale}),(\ref{eq:rescale_g})
and (\ref{eq:scaled}) in the continuum, and of the corresponding
result in the lattice theory of gravity, Eq.~(\ref{eq:rescale_latt}).
As a result of this conflict, it seems now possible to find spurious
gauge-dependent separate renormalization group trajectories for 
$G(k)$ and $\lambda(k)$, in disagreement with most of the arguments 
presented previously in this work, 
including the explicit gauge-independence of the perturbative result of
Eq.~(\ref{eq:g-ren}).
In light of these issues, it would seem that the RG trajectory
for the {\it dimensionless} combination $ G(k) \lambda (k)$ should be
regarded as more trustworthy.
The second problem is that the running of $\lambda (k)$ claimed
in this approach seems largely accidental, presumably due
to the diffeomorphism violating cutoff.
The latter allows such a running, in spite of the fact that, as we 
have shown earlier, it is inconsistent with general covariance.
One additional and somewhat unrelated problem is the fact that the
above method, at least in its present implementation, is essentially 
perturbative and still relies on the weak field expansion. 
It is therefore unclear how such a method could possibly give rise
to an explicit nonperturbative correlation length $\xi$
[see Eq.~(\ref{eq:m-cont})], which after all is non-analytic in $G$.

\vskip 30pt

\section{Conclusions}

\label{sec:conc}

In this paper we have examined the issue of whether the cosmological
constant of quantized gravitation can run with scale.
The relevance of this problem arises at a fundamental level, but
has also possible implications for observational cosmology,
where a scale dependence of $\lambda$ in the form
of a $\lambda (a(t))$ is sometimes assumed. 
We have examined this issue from a variety of viewpoints, which
included the continuum and lattice formulations for the 
gravitational path integral, with various scaling properties that come
with it; the perturbative treatment of gravity; and finally 
from insights gained through the formulation
of manifestly covariant effective actions and effective field equations.

The key message seems that the cosmological constant cannot run with
scale, if general covariance is preserved.
Instead, evidence from the nonperturbative path integral treatment of
quantum gravity points to the fact that the observed 
effective long-distance cosmological is a renormalization
group invariant quantity, related to the fundamental RG scale
$\xi$, and thus to a vacuum condensate of the gravitational field.
In analogy to the corresponding scale for non-Abelian gauge theories, 
$\xi$ cannot run, and represents instead a dynamically generated,
nonperturbative mass-like parameter.
That this is possible is a highly non-trivial result of the renormalization
group treatment, of the Callan-Symanzik RG equations for $G$, and of the phase
structure of four-dimensional gravity.

In closing, let us pursue here again what appears as a deep analogy 
between gravity on the one hand, and gauge theories and magnets on the
other.
First write down the three field equations
for gravity, quantum electrodynamics (made massive via the Higgs mechanism)
and a scalar field.
They read 
\bea
R_{\mu\nu} \, - \, \half \, g_{\mu\nu} \, R \, + \, \lambda \, g_{\mu\nu}
& = & 8 \pi \, G \; T_{\mu\nu}
\nonumber \\
\partial^\mu \, F_{\mu\nu} \, + \, \mu^2 \, A_\nu
& = & 4 \pi \, e \, j_{\nu}
\nonumber \\
\partial^\mu \partial_\mu \, \phi \, + \, m^2 \, \phi
& = & { g \over 3! } \, \phi^3  \; ,
\label{eq:masses}
\eea
and are used here to represent the field equations relevant 
for a boson-mediated long range force.
Now, all three mass-like parameters on the left
($\lambda$, $\mu$ and $m$) are considered RG invariants (this
is well known for the last two cases), whereas
all three couplings of the r.h.s. are known to be scale-dependent.
Furthermore, in all three cases the relevant renormalized mass
parameter is related to the fundamental correlation length, $m = 1 / \xi $.
More generally, in non-Abelian gauge theories the nonperturbative
mass parameter (sometimes referred to as the mass gap) is also an RG invariant;
that such a mass scale can be generated dynamically is a non-trivial
result of the renormalization group.

Here we want to point out that there seems to be a fundamental 
relationship between the nonperturbative scale $\xi$
(or inverse renormalized mass) and a non-vanishing vacuum
condensate for the three theories,
\beq
\langle \, R \, \rangle \;  \simeq \;  { 1 \over \xi^2 }
\;\;\;\;\;\;\;\;
\langle \, F_{\mu\nu}^2 \, \rangle  \;  \simeq \;  { 1 \over \xi^4 }
\;\;\;\;\;\;\;\;
\langle \, \phi \, \rangle \;  \simeq \;  { 1 \over \xi } \; .
\label{eq:vevs}
\eeq
In all three cases the vacuum condensate's dependence
on the correlation length $\xi$ is fixed by the 
mass dimension of the field appearing in it.
In the gauge theory case, this is due to the vanishing relevant anomalous
dimension, which in turn follows from current conservation.
One more notable example that comes to mind is the fermion condensate in non-Abelian
gauge theories, $ \langle \, \bar \psi \psi \, \rangle \, \simeq \, 1 / \xi^3 $.
The last result listed in Eq.~(\ref{eq:vevs}), for a scalar field with 
a non-vanishing vacuum expectation value, is the field theory analog
of what happens in a ferromagnet. 
There in the magnetized phase, $T < T_c $, the general result in $d$
dimensions is 
$ \langle \, \phi \, \rangle \, \simeq \, 1 / \xi^{\beta  / \nu } $ 
close to the critical point, where $\nu$ and $\beta$ are some
exponents;
then already for Ising spins in four dimensions one has 
$ \langle \, S \, \rangle \simeq 1 / \xi $, given the exponents 
$\nu = \beta = \half $ in $d=4$.
So, in the end, the relationship between the fundamental
nonperturbative correlation length $\xi$ and the vacuum condensate 
starts to look a lot less exotic than what might have seemed at first sight.

\vspace{10pt}

Note added: After this work was submitted, we were informed that similar
conclusions had been reached by the authors of \cite{foo08}, by
considering the effective action for a self-interacting scalar field
in a classical gravitational background.
A closely related paper also dealt with the issue of
a possible running of $\lambda (k) $ \cite{sha09}.


\vspace{10pt}

{\bf Acknowledgements}

One of the authors (H.W.H.) is grateful to Gabriele Veneziano
for discussions that largely motivated the present study.
The work of H.W.H. was supported in part by the Max 
Planck Gesellschaft zur F\" orderung der Wissenschaften, and
by the University of California.
He wishes to thank Hermann Nicolai and the
Max Planck Institut f\" ur Gravitationsphysik (Albert-Einstein-Institut)
in Potsdam for warm hospitality. 
The work of R.T. was supported in part by a DED GAANN Student Fellowship.


\vfill


\end{document}